\documentclass[3p,,preprint,12pt]{elsarticle}
\makeatletter\if@twocolumn\PassOptionsToPackage{switch}{lineno}\else\fi\makeatother

      \makeatletter
\usepackage{wrapfig}
\newcounter{aubio}

\usepackage{tabularx}

\long\def\@@bioItem#1#2{
 \stepcounter{aubio}
 \expandafter\gdef\csname authorName\theaubio\endcsname{#1}
 \expandafter\gdef\csname authorDetails\theaubio\endcsname{#2}
}

\newcommand{\checkheight}[1]{%
  \par \penalty-100\begingroup%
  \setbox8=\hbox{#1}%
  \setlength{\dimen@}{\ht8}%
  \dimen@ii\pagegoal \advance\dimen@ii-\pagetotal
  \ifdim \dimen@>\dimen@ii
    \break
  \fi\endgroup}

\makeatother

\usepackage{tabulary,xcolor}
\usepackage{amsfonts,amsmath,amssymb}
\usepackage[T1]{fontenc}
\makeatletter
\let\save@ps@pprintTitle\ps@pprintTitle
\def\ps@pprintTitle{\save@ps@pprintTitle\gdef\@oddfoot{\footnotesize\itshape \null\hfill\today}}
\def\hlinewd#1{%
  \noalign{\ifnum0=`}\fi\hrule \@height #1%
  \futurelet\reserved@a\@xhline}

\AtBeginDocument{\ifNAT@numbers \biboptions{sort&compress}\fi}

\makeatother

\usepackage{ifluatex}
\ifluatex
\usepackage{fontspec}
\defaultfontfeatures{Ligatures=TeX}
\usepackage[]{unicode-math}
\unimathsetup{math-style=TeX}
\else 
\usepackage[utf8]{inputenc}
\fi 
\ifluatex\else\usepackage{stmaryrd}\fi

\usepackage{url,multirow,morefloats,floatflt,cancel,tfrupee}
\makeatletter

\AtBeginDocument{\@ifpackageloaded{textcomp}{}{\usepackage{textcomp}}}
\makeatother
\usepackage{colortbl}
\usepackage{xcolor}
\usepackage{pifont}
\usepackage[nointegrals]{wasysym}
\usepackage{float}
\makeatletter

\usepackage{booktabs}
\usepackage{tabularx}
\usepackage{tcolorbox}

\def\mcWidth#1{\csname TY@F#1\endcsname+\tabcolsep}

\def\cAlignHack{\rightskip\@flushglue\leftskip\@flushglue\parindent\z@\parfillskip\z@skip}
\def\rAlignHack{\rightskip\z@skip\leftskip\@flushglue \parindent\z@\parfillskip\z@skip}

\@ifundefined{etal}{}{}

\usepackage{ifxetex}
\ifxetex\else\if@twocolumn\@ifpackageloaded{stfloats}{}{\usepackage{dblfloatfix}}\fi\fi

\AtBeginDocument{
\expandafter\ifx\csname eqalign\endcsname\relax
\def\eqalign#1{\null\vcenter{\def\\{\cr}\openup\jot\m@th
  \ialign{\strut$\displaystyle{##}$\hfil&$\displaystyle{{}##}$\hfil
      \crcr#1\crcr}}\,}
\fi
}

\AtBeginDocument{%
  \@ifpackageloaded{endfloat}%
   {\renewcommand\efloat@iwrite[1]{\immediate\expandafter\protected@write\csname efloat@post#1\endcsname{}}}{\newif\ifefloat@tables}%
}%

\def\BreakURLText#1{\@tfor\brk@tempa:=#1\do{\brk@tempa\hskip0pt}}
\let\lt=<
\let\gt=>
\def\processVert{\ifmmode|\else\textbar\fi}

\@ifundefined{subparagraph}{
\def\subparagraph{\@startsection{paragraph}{5}{2\parindent}{0ex plus 0.1ex minus 0.1ex}%
{0ex}{\normalfont\small\itshape}}%
}{}

\newcommand\role[1]{\unskip}
\newcommand\aucollab[1]{\unskip}
  
\@ifundefined{tsGraphicsScaleX}{\gdef\tsGraphicsScaleX{1}}{}
\@ifundefined{tsGraphicsScaleY}{\gdef\tsGraphicsScaleY{.9}}{}
\def\checkGraphicsWidth{\ifdim\Gin@nat@width>\linewidth
	\tsGraphicsScaleX\linewidth\else\Gin@nat@width\fi}

\def\checkGraphicsHeight{\ifdim\Gin@nat@height>.9\textheight
	\tsGraphicsScaleY\textheight\else\Gin@nat@height\fi}

\def\fixFloatSize#1{}%\@ifundefined{processdelayedfloats}{\setbox0=\hbox{\includegraphics{#1}}\ifnum\wd0<\columnwidth\relax\renewenvironment{figure*}{\begin{figure}}{\end{figure}}\fi}{}}
\let\ts@includegraphics\includegraphics

\def\inlinegraphic[#1]#2{{\edef\@tempa{#1}\edef\baseline@shift{\ifx\@tempa\@empty0\else#1\fi}\edef\tempZ{\the\numexpr(\numexpr(\baseline@shift*\f@size/100))}\protect\raisebox{\tempZ pt}{\ts@includegraphics{#2}}}}

\DeclareMathAlphabet{\mathpzc}{OT1}{pzc}{m}{it}

\def\URL#1#2{\@ifundefined{href}{#2}{\href{#1}{#2}}}

\def\UrlOrds{\do\*\do\-\do\~\do\'\do\"\do\-}%
\g@addto@macro{\UrlBreaks}{\UrlOrds}

\edef\fntEncoding{\f@encoding}

\makeatother

\newif\ifmultipleabstract\multipleabstractfalse%
    \makeatletter
\gdef\emailauthor#1#2{\stepcounter{ead}
      \g@addto@macro\@elseads{\raggedright
      \let\corref\@gobble
      \eadsep\texttt{#1} (#2)
      \def\eadsep{\unskip,\space}}
}

\makeatother
  
\let\citep\cite
\let\citet\cite    
  
\usepackage{float}

\begin{document}

\begin{frontmatter}
    \title{
  AI Infrastructure Sovereignty}
    
\author[ciena]{Sergio Cruzes\fnref{disclaimer}}

\address[ciena]{Optical Network Engineering, Ciena Brazil,
Ciena, Av. das Na\c{c}\~{o}es Unidas, 14.171 -- 15\textordmasculine{}
andar -- Marble Tower -- Salas 1563/1564,
S\~{a}o Paulo, 04794-000, SP, Brazil}

\fntext[disclaimer]{The views, analyses, and conclusions expressed
in this paper are those of the author alone and do not represent
the position, policy, or endorsement of Ciena Corporation or any
of its affiliates. This research was conducted by the author
independently in a personal academic capacity.}
  
%\maketitle
\begin{abstract}
Artificial intelligence is no longer just a software problem. It has become an infrastructure problem. Training and inference at scale now depend on tightly connected data centers, high-capacity optical networks, and energy systems operating close to their physical and environmental limits. In this context, control over data and algorithms is not enough. Real AI sovereignty depends on the ability to deploy, operate, and adapt infrastructure under constraints such as energy availability, sustainability requirements, and network reach.

This tutorial-survey introduces the concept of AI infrastructure sovereignty, defined as the ability of a region, operator, or nation to maintain operational control over AI systems within these constraints. The central idea is that sovereignty emerges from the joint design of three layers: AI-oriented data centers, optical transport networks, and control frameworks that provide real-time visibility and coordination across them.

We first examine how AI workloads are reshaping data center design, pushing power densities higher, increasing cooling demands, and tightening the relationship with local energy systems. In this setting, factors such as carbon intensity and water usage become hard limits on where and how AI can be deployed. We then look at optical networks as the backbone of distributed AI, showing how latency, capacity, failure domains, and jurisdictional boundaries directly influence what can be achieved in practice.

Building on this foundation, the paper highlights the role of telemetry, agentic AI, and digital twins as key enablers of operational sovereignty. Together, they make it possible to monitor, coordinate, and validate system behavior across compute, network, and energy domains in a closed loop.

The tutorial concludes with a reference architecture for sovereign AI infrastructure that integrates telemetry pipelines, agent-based control, and digital twins, treating sustainability as a core design constraint rather than an afterthought.
\end{abstract}

    
  \end{frontmatter}

\section{Introduction}
\label{sec:intro}

Artificial intelligence has become a defining force in the evolution of digital infrastructure. What once progressed mainly through advances in algorithms and software is now increasingly shaped by physical constraints: power availability, cooling capacity, water usage, optical connectivity, and integration with the electrical grid. Large-scale AI systems already operate at the scale of industrial facilities, with data centers drawing tens to hundreds of megawatts per site and exhibiting rapid, high-amplitude power fluctuations that challenge both facility design and grid stability~\cite{Patterson2021Carbon,IEA2023Datacenters,Chen2025Grid}. As a result, the limits on AI progress are no longer primarily computational, but infrastructural.

Modern AI workloads, especially large language models, rely on tightly coordinated computation across dense clusters of accelerators, supported by high-capacity data centers and low-latency optical networks. Training concentrates extreme power and cooling demand over sustained periods, while inference runs continuously at global scale, often dominating the overall energy and water footprint~\cite{Hoxha2025LLM,Shehabi2024united}. These characteristics tightly couple compute, network, and energy systems, making performance, resilience, and sustainability inseparable from how infrastructure is designed and operated.

\subsection{From software sovereignty to infrastructure sovereignty}

These realities call for a shift in how AI sovereignty is understood. Most existing discussions focus on data localization, jurisdiction, privacy, and model ownership. Frameworks such as GDPR~\cite{GDPR2016} and the CLOUD Act~\cite{CloudAct2018} show how sovereignty has been approached through legal and regulatory mechanisms. Intellectual property protections over data, models, and software add another layer of formal control. While these remain important, they do not fully explain who can actually deploy, operate, and evolve AI systems at scale~\cite{FutureCIO2025,Fratini2024DigitalSovereignty}.

In practice, sovereignty increasingly depends on access to physical resources such as electricity, water, land, fiber infrastructure, and automation capabilities, along with the ability to coordinate them under real operational constraints.

This tutorial draws a key distinction between \emph{legal sovereignty} and \emph{operational sovereignty}. Legal sovereignty concerns jurisdiction, compliance, and formal ownership, as reflected in regulatory frameworks and IP protections. Operational sovereignty, by contrast, is about what can actually be done: the ability to observe system state, make decisions based on local conditions, validate those decisions, and act within defined policies and physical limits. It depends on visibility, control authority, and the ability to execute reliably across compute, network, and energy systems. In this sense, telemetry, automation, and closed-loop control are not just operational tools, but core enablers of sovereign AI infrastructure.

Sustainability adds another dimension to this picture. It is not just a reporting requirement, but a real constraint that must be enforced in operation. Metrics such as Power Usage Effectiveness (PUE)~\cite{TheGreenGrid2007}, which measures the overhead of cooling and power delivery relative to IT load, become critical. As rack power densities increase, even small inefficiencies translate into large absolute energy waste, with direct implications for carbon emissions and water consumption.

\begin{figure}[t]
  \centering
  \includegraphics[width=\linewidth]{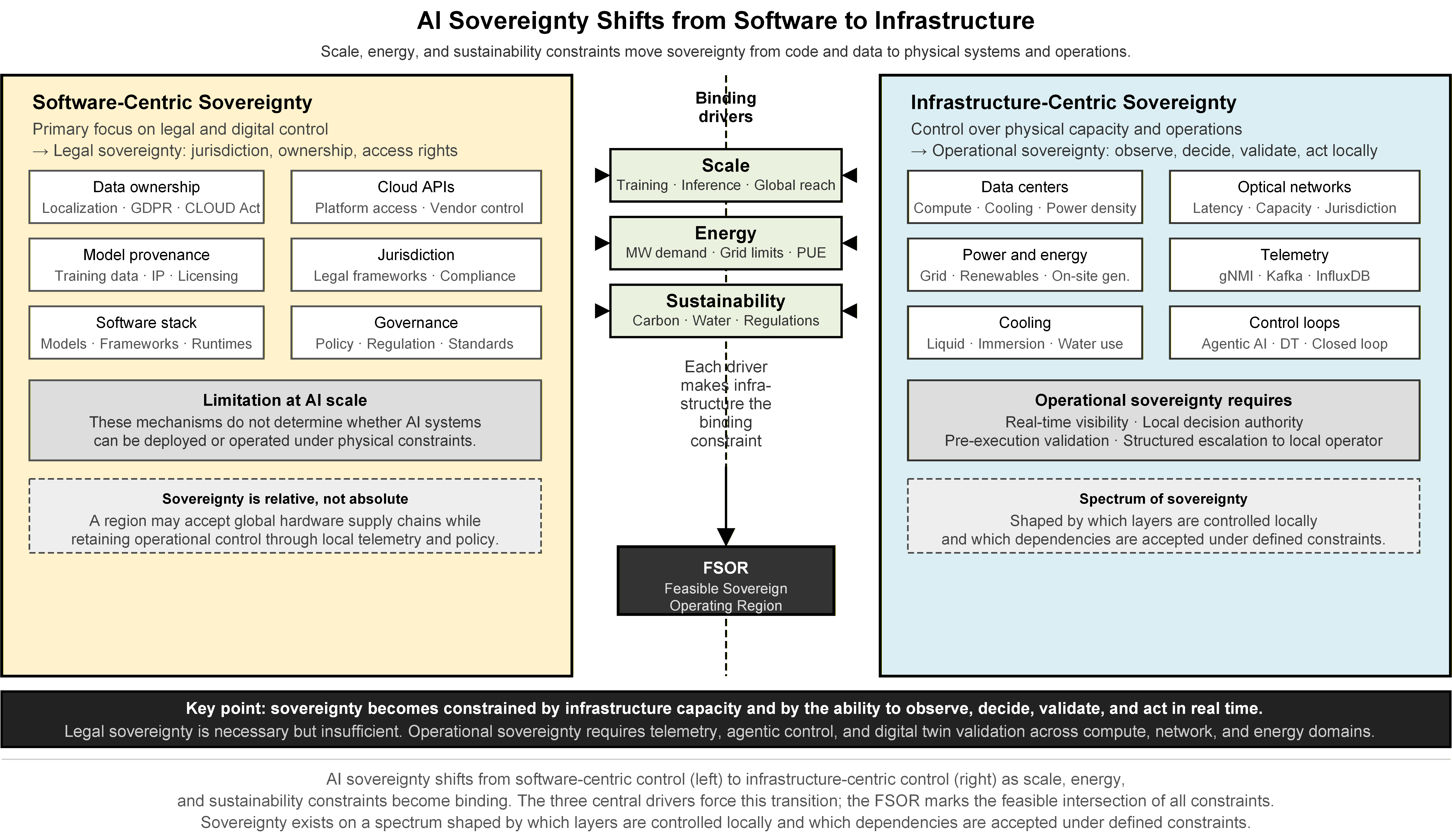}
  \caption{%
    \textbf{AI sovereignty shifts from software-centric to
    infrastructure-centric control as scale, energy, and
    sustainability constraints become binding.}
    The left panel shows software-centric sovereignty ---
    legal control over data (GDPR, CLOUD Act), model IP,
    jurisdiction, and governance --- which is necessary but
    insufficient at AI scale.
    The three central drivers (scale at MW-class power demand,
    energy including grid limits and PUE, and sustainability
    including carbon, water, and regulations) force a
    transition: each makes physical infrastructure the binding
    constraint on AI deployment and operation, independently
    of software capability or regulatory compliance.
    Their intersection defines the Feasible Sovereign Operating
    Region (FSOR) --- the space within which all three
    constraints can be jointly satisfied.
    The right panel shows infrastructure-centric sovereignty
    --- operational control over data centers, optical
    networks, power and energy systems, telemetry, cooling,
    and closed-loop control systems (agentic AI, digital twins,
    and closed-loop automation) --- which is the practical
    expression of operational sovereignty.
    Sovereignty exists on a spectrum shaped by which layers
    are controlled locally and which dependencies are accepted
    under defined constraints~\cite{Fratini2024DigitalSovereignty,
    MITTechReview2026Sovereignty}.%
  }
  \label{fig:Sovereignty_shift}
\end{figure}

Figure~\ref{fig:Sovereignty_shift} captures this transition. 
Traditional views of sovereignty are centered on software and legal control: data ownership, model provenance, intellectual property (IP), and regulatory compliance. In this context, IP refers to the legal ownership of training data, model weights, algorithms, and software stacks, including the rights to use, license, and restrict access to them. These mechanisms define who formally controls AI assets, but they do not by themselves determine whether AI systems can be deployed or operated at scale.
An infrastructure-centric perspective shifts the focus to physical and operational constraints, including power capacity, cooling efficiency, optical connectivity, and real-time system visibility. Three factors drive this transition: scale, energy, and sustainability. Each imposes limits on where and how AI systems can operate, and together they define what this tutorial calls the Feasible Sovereign Operating Region (FSOR), introduced in Section~\ref{sec:reference_architecture}. 
Sovereignty, in this sense, is not absolute. It exists on a spectrum shaped by which parts of the stack are controlled locally and which dependencies are accepted under clearly defined constraints~\cite{Fratini2024DigitalSovereignty}.

\subsection{Infrastructure as the binding constraint}

Recent geopolitical developments reinforce this view. Export controls, sanctions, and concentration of platforms have shown that access to compute and infrastructure can be restricted by external decisions~\cite{Esposito2025}. As AI becomes central to economic activity and national security, many regions are seeking ways to reduce this exposure and achieve greater autonomy~\cite{MITTechReview2026Sovereignty}.

The physical demands of AI workloads make infrastructure the key constraint. Training clusters can require hundreds of megawatts, generate extreme heat densities, and introduce rapid power fluctuations that challenge both facilities and grid interfaces~\cite{Patterson2021Carbon,Chen2025Grid}. Inference workloads run continuously at large scale, increasing total energy, water, and network demand over time~\cite{Hoxha2025LLM,ODonnellCrownhart2025}. In this context, data centers become strategic assets, and optical networks become critical control surfaces that define latency, reach, and resilience~\cite{korde2025sovereign,Shehabi2024united}.

Regions with limited grid capacity, high carbon intensity, water constraints, or insufficient connectivity face real barriers to deploying AI systems, regardless of their software capabilities~\cite{IEA2023Datacenters,Chen2025Grid,Xiao2025}. On the other hand, regions that can align clean energy, efficient cooling, and high-capacity optical transport gain a lasting advantage. This makes it clear that models of sovereignty based only on data or software assume an infrastructure that is effectively unlimited and interchangeable. That assumption no longer holds. Compute cannot be moved freely without considering power, cooling, latency, and reliability constraints~\cite{korde2025sovereign,AIBottleneck2026}.

AI infrastructure sovereignty can therefore be understood as the ability to design, operate, and control AI systems within these constraints, without excessive reliance on external actors~\cite{IBM2026}. In practical terms, it comes down to operational control: being able to observe the system, make decisions, validate them, and act within local limits and policies.

\subsection{Telemetry and agentic AI as the operational layer}

Achieving this level of control requires mechanisms that connect high-level intent to real infrastructure behavior. Telemetry provides continuous visibility into the state of data centers, networks, and energy systems~\cite{Cruzes2026,Clemm2015RFC7575}. Agentic AI builds on this by making decisions, coordinating actions across domains, and enforcing policies under uncertainty~\cite{Cruzes2026,Kiasari2026AgenticGrid}. Digital twins provide a way to simulate and validate actions before they are applied, reducing risk in complex systems~\cite{Cruzes2026}. Large language models help translate human intent into structured objectives and explain system behavior in accessible terms~\cite{Cruzes2025llm}.

Together, these elements form a closed-loop control system. Telemetry enables observation, agentic systems enable decision-making, digital twins provide validation, and execution systems carry out actions within defined constraints. This approach extends earlier work in network and data center automation to a broader setting that includes energy systems and sustainability constraints, forming a unified control framework for sovereign AI infrastructure.

\subsection{Scope and organization}

This tutorial provides a cross-layer view of AI infrastructure sovereignty grounded in engineering practice. It is intended as a system-level synthesis for researchers, operators, and policymakers. It brings together two areas that have often been treated separately: discussions of AI sovereignty focused on data and regulation, and work on AI infrastructure focused on scaling and efficiency. The central argument is that sovereignty in the AI era depends on the ability to observe, control, and optimize physical infrastructure under real constraints.

The rest of the paper is organized as follows. Section~2 shows how AI workloads translate into physical infrastructure constraints. Section~3 examines data centers and the sustainability limits that shape deployment. Section~4 analyzes optical networks and their role in defining latency, capacity, and control boundaries. Sections~5 and~6 introduce telemetry, agentic AI, and digital twins as the operational foundation. Section~7 presents a reference architecture, and Section~8 discusses sustainability-aware automation and its policy implications.

\section{From AI Models to Physical Infrastructure}
\label{sec:ai_to_infra}

AI infrastructure cannot be understood as a purely digital stack. At
scale, it is shaped by a hierarchy of physical constraints that start
with power availability and extend upward through cooling, networking,
and operations. Higher layers can improve efficiency and utilization,
but they cannot bypass limits imposed by electrical capacity, heat
dissipation, or physical space. As AI systems grow, infrastructure
design becomes less about abstraction and more about managing these
constraints directly. This shift is central to sovereignty: control
over models and data is necessary, but it does not matter if the
underlying systems cannot be built, powered, or cooled.

\subsection{The infrastructure demands of modern AI workloads}

AI progress is often described in terms of model size, data, and
algorithmic advances. But behind that narrative is a deeper change in
how workloads behave. Modern AI systems, especially large-scale
foundation models, rely on tightly synchronized computation across
thousands of accelerators acting as a single system~\cite{Bommasani2021,
Patterson2021Carbon}. Training concentrates extreme demand into short,
intense periods, while inference spreads continuous demand across
global platforms~\cite{Hoxha2025LLM,Shehabi2024united}. Together, these
patterns make AI far more dependent on physical infrastructure than
traditional software workloads~\cite{Arora2025}.

\begin{figure}[t]
\centering
\includegraphics[width=\linewidth]{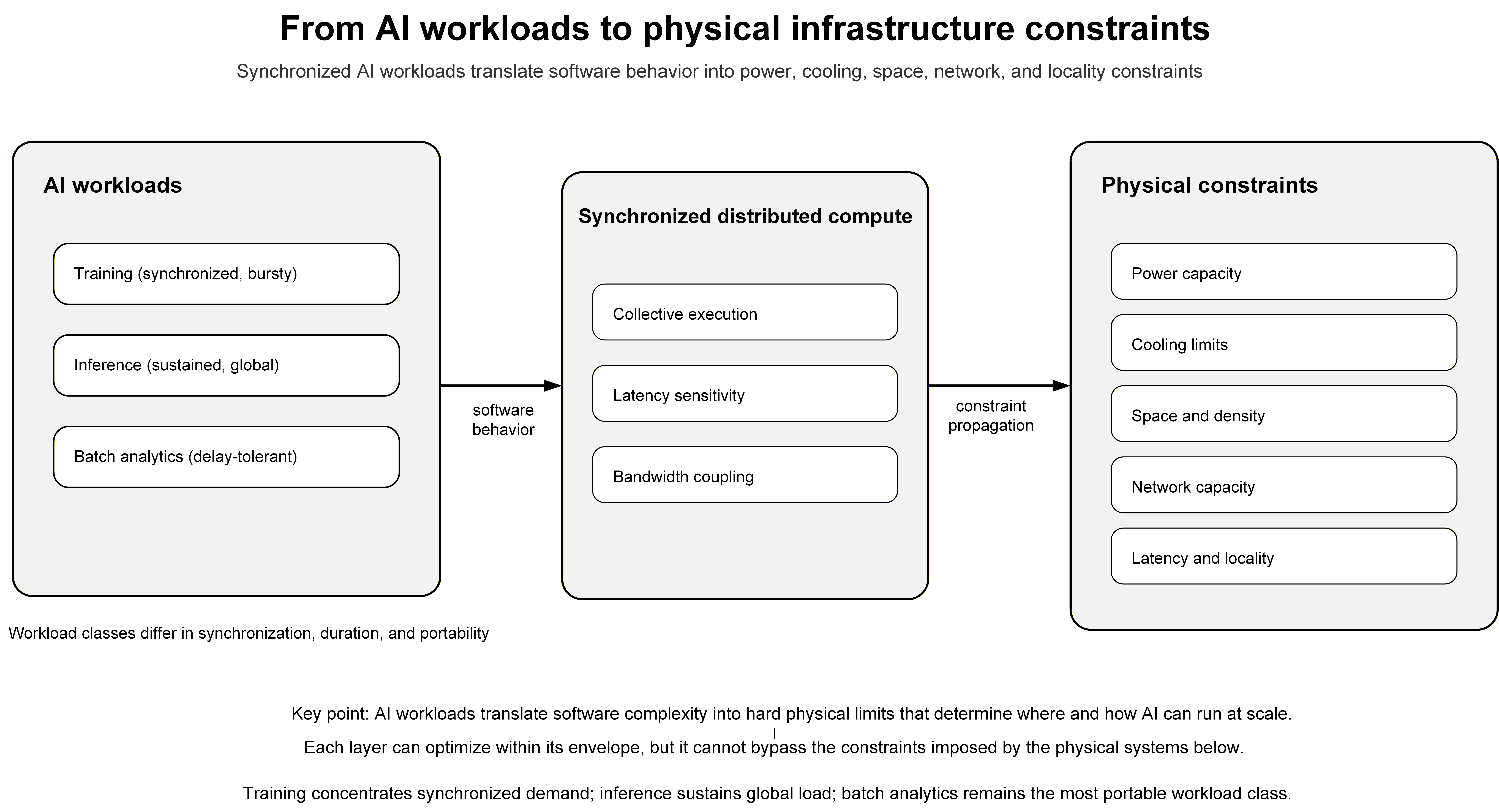}
\caption{From AI workloads to physical infrastructure constraints.
  Synchronized training and large-scale inference translate software
  demands into hard limits on power delivery, cooling capacity, physical
  space, and network bandwidth. Each layer can optimize within its
  envelope but cannot bypass the constraints imposed by the layer
  below.}
\label{fig:models_to_infra}
\end{figure}

Figure~\ref{fig:models_to_infra} illustrates how this transition
happens. Software-level characteristics, such as synchronization,
latency sensitivity, and bandwidth requirements, propagate downward
into concrete infrastructure demands. What begins as a property of the
workload becomes a requirement for power delivery, cooling capacity,
space, and network bandwidth. In other words, workload behavior turns
into physical constraint.

\subsection{Power, cooling, and space as primary constraints}

Once compute is tightly synchronized, pressure builds across all
supporting systems at the same time. Power demand is not only high,
but highly dynamic. Rapid fluctuations driven by collective
communication patterns challenge conventional power distribution and
backup systems~\cite{choukse2025powerstab,Chen2025Grid}. Cooling must
handle extreme heat densities, often beyond what air-based systems can
manage, which is why liquid cooling and immersion solutions are
becoming standard~\cite{Shehabi2024united,IEA2023Datacenters}. Physical
space adds another constraint: dense racks, power equipment, and
cooling systems compete for limited floor area.

At this point, the limiting factor is no longer just the availability
of accelerators, but whether a site can support them. Grid capacity,
transformers, cooling infrastructure, and water availability all place
hard limits on how much AI capacity can be deployed at a given
location~\cite{IEA2023Datacenters,Chen2025Grid}. These limits vary
widely across regions, creating uneven conditions for AI deployment
that software alone cannot overcome.

This marks a shift from earlier cloud models, where workloads could be
moved relatively freely between data centers. AI workloads are far less
portable. They tie computation to locations that can meet strict power,
cooling, and networking requirements, exposing operators directly to
local infrastructure constraints~\cite{Shehabi2024united}. Regions with
reliable energy, efficient cooling, and strong optical connectivity can
support large-scale AI systems and expand over time~\cite{IEA2023Datacenters}.
Others face structural barriers that cannot be solved through software
optimization. Even when hardware is available, a lack of power or
cooling makes it unusable.

\subsection{Workload classes and their infrastructure signatures}

Not all AI workloads place the same demands on infrastructure.
Table~\ref{tab:ai_workload_infra} compares three main classes:
training, inference, and batch analytics. This distinction matters for
both sovereignty and sustainability planning. Training workloads are
tightly coupled and must run within strict latency bounds, making them
hard to relocate. Inference workloads are more distributed, but still
need to remain close to users to meet latency requirements. Batch
analytics workloads are more flexible and can be shifted in time or
location, making them the main candidates for adapting to energy or
sustainability constraints.

\begin{table}[t]
\centering
\caption{AI workload classes and their infrastructure signatures.
  Batch analytics workloads are delay-tolerant and the primary
  candidates for sustainability-driven temporal and spatial shifting.
  Training workloads are spatially constrained by synchronization
  requirements and cannot be shifted without violating collective
  communication latency bounds.}
\label{tab:ai_workload_infra}
\begin{tabular}{p{2.6cm} p{2.7cm} p{3.0cm} p{3.1cm} p{2.7cm}}
\hline
AI workload & Power profile & Cooling stress & Network demand
  & Portability \\
\hline
Training        & Bursty, high          & Extreme
  & Very high     & Low    \\
\hline
Inference       & Sustained             & Moderate--high
  & High          & Medium \\
\hline
Batch analytics & Variable, schedulable & Moderate
  & Low--moderate & High   \\
\hline
\end{tabular}
\smallskip

\noindent\footnotesize Entries are qualitative relative comparisons
within each column rather than absolute measurements; absolute values
depend on cluster size, hardware generation, and deployment
configuration.
\end{table}

In practical terms, portability refers to how easily a workload can be
moved without breaking its performance constraints. Training has very
low portability because of strict synchronization requirements.
Inference has moderate portability, since it can be replicated but must
stay close to demand. Batch workloads are highly portable and can be
shifted to optimize for energy availability or carbon intensity. From
a sovereignty perspective, this matters: only portable workloads can be
moved to adapt to constraints. Non-portable workloads require local
infrastructure to already be in place.

\subsection{Infrastructure as a sovereignty boundary}

This hierarchy of physical constraints has a direct impact on
sovereignty. Control over AI systems is not determined only by data or
models, but by who can provide power, cooling, connectivity, and
operational control. Regions without sufficient grid capacity, water,
or fiber infrastructure face limits that are physical rather than
legal~\cite{Toni2025,IEA2023Datacenters}.

Sovereignty therefore emerges unevenly across infrastructure layers.
The workload distinctions in Table~\ref{tab:ai_workload_infra} make this
clear. A region might support batch workloads but not large-scale
training, meaning it has only partial AI sovereignty. Closing this gap
requires investment not in software, but in energy systems, cooling,
and optical networks. The next sections examine these constraint
domains in more detail, starting with data center sustainability limits.

\section{AI-Oriented Data Centers and Sustainability Limits}
\label{sec:datacenters}
% ============================================================

One of the central challenges in AI infrastructure planning is the mismatch
between how quickly digital systems can scale and how slowly physical
infrastructure evolves. Compute capacity can be deployed relatively
fast, but power grids, substations, cooling systems, and permitting
processes move on much longer timelines. Because of this mismatch,
capital alone does not determine where AI can be built. Physical
infrastructure becomes the limiting factor, and sustainability
constraints, such as energy availability, carbon intensity, and water
usage, are often where those limits show up most clearly.

\subsection{How AI workloads reshape data center design}

AI-oriented data centers are fundamentally different from traditional
enterprise or cloud facilities. Large training and inference clusters
pack a high concentration of accelerators into a single space, pushing
rack power densities far beyond historical norms. This shift forces a
move toward high-density layouts and requires a complete rethink of
mechanical and electrical systems~\cite{Shehabi2024united,Buyya2013EnergyEfficient}.

Traditional data centers were designed for moderate and relatively
stable heat loads. AI systems are not. They generate sustained,
high-intensity heat and rapid load fluctuations driven by tightly
synchronized computation. As a result, air-based cooling, while still
used in hybrid setups, quickly reaches its limits. Once rack densities
move beyond roughly 20--30\,kW, which is now common in GPU clusters,
air cooling alone cannot keep up without introducing impractical
airflow requirements. This is why liquid cooling, both direct-to-chip
and immersion, is becoming standard. These approaches remove heat
closer to the source and provide more stable thermal conditions under
continuous AI workloads~\cite{IEA2023Datacenters,Shehabi2024united}.

What makes this shift particularly important is how tightly different
systems become coupled. Power, cooling, and water usage are no longer
independent design choices. Increasing electrical density raises heat,
which determines the cooling approach, which in turn affects water
consumption. A decision driven by compute demand can end up being
limited by a water permit. Once air cooling is no longer viable,
design moves from optimizing individual components to balancing
trade-offs across the entire system. This is why sustainability
constraints need to be considered together, not in isolation.

Figure~\ref{fig:sustainability_triangle} captures this relationship.
It shows how energy availability, carbon limits, and water and cooling
feasibility jointly define the space in which AI deployment is
possible. The feasible region, where all three constraints are
satisfied, is not fixed. Over time, it is expected to shrink as water
availability becomes more constrained~\cite{IPCC2021WG1}, carbon
regulations tighten~\cite{Xiao2025,IEA2023Datacenters}, and grid
expansion struggles to keep pace with demand~\cite{IEA2023Datacenters}.
The boundaries shown in the figure are directional rather than
precise, since the rate of change varies by region, grid mix, and
local conditions. The implication is straightforward: a site that is
viable today may not remain viable in the future without investment in
clean energy, efficient cooling, or local generation.

\subsection{Energy: the dominant constraint}

Among all constraints, energy is the most immediate and often the most
limiting. AI data centers require large, continuous power supplies,
but they also need to handle rapid fluctuations caused by synchronized
workloads~\cite{choukse2025powerstab,Chen2025Grid}. Grid
interconnection capacity, transformer limits, and transmission
constraints often determine how large a facility can be, even before
any hardware is installed~\cite{Shehabi2024united}. In many regions,
long permitting cycles and slow grid upgrades further restrict how
quickly capacity can expand.

There is also an important dynamic aspect. AI workloads do not draw
power evenly. Collective operations during training can create sharp,
short-lived spikes that exceed average consumption by a wide margin.
This means that infrastructure must be sized for peak demand, not
average usage. As a result, part of the available grid capacity cannot
be fully utilized, increasing costs and reducing efficiency.
Facilities that underestimate this behavior risk instability or
failure during sustained workloads.

\subsection{Carbon intensity: from preference to constraint}

Carbon intensity has moved from being a sustainability preference to
a real constraint on where AI infrastructure can be deployed.
Operators are no longer concerned only with how much energy is
available, but also with how that energy is produced. Regions with
low-carbon grids or access to renewable generation are increasingly
favored~\cite{Cote2025,Nutt2025}. Studies show that identical systems
can have very different environmental impacts depending solely on the
local energy mix~\cite{Xiao2025}.

In regions where energy comes primarily from fossil sources, AI
expansion faces growing regulatory and financial pressure. Carbon
reporting requirements and, in some cases, operational limits are
becoming more common~\cite{IEA2023Datacenters}. This turns carbon
intensity into a site-level filter. Even if power and hardware are
available, a site may not be eligible to host certain workloads if its
emissions exceed defined thresholds. From a sovereignty perspective,
this creates a new kind of barrier that cannot be solved with compute
investment alone. It requires changes in energy sourcing or access to
clean generation.

\subsection{Water: the underestimated constraint}

Water is often overlooked in AI infrastructure planning, but it can be
just as limiting as energy. High-density cooling systems, especially
those using evaporative methods, can consume large volumes of water
per unit of compute~\cite{YanezBarnuevo2025}. Local climate plays a
major role here. Temperature, humidity, and seasonal variation all
affect cooling efficiency and water demand~\cite{IEA2023Datacenters}.

This introduces a strong seasonal effect. A facility that operates
comfortably within its limits in winter may exceed its water permit
during summer, when both cooling demand and water stress increase.
In regions already facing water scarcity, this can make large AI
deployments impractical, even when sufficient electrical capacity is
available~\cite{Xiao2025}. Unlike carbon, which can sometimes be
managed through scheduling, water limits are typically fixed at the
facility level and are harder to work around. Climate projections
suggest that these constraints will become more severe over time,
with longer and more intense dry periods~\cite{IPCC2021WG1}.

\subsection{Sustainability as a sovereignty boundary}

Taken together, energy, carbon, and water define hard boundaries for
AI deployment~\cite{IEA2023Datacenters,Xiao2025}. These limits cannot
be fully offset through incremental improvements in hardware or
software. Once they are reached, further expansion requires changes in
infrastructure, not optimization.

This has direct implications for sovereignty. Control over AI is not
just about data, models, or regulation. It depends on whether a region
can provide the energy, cooling, and infrastructure needed to run
these systems. A deployment may satisfy all legal requirements and
still be infeasible because of physical constraints. In those cases,
sovereignty is limited by infrastructure, not policy.

At the same time, regions that can combine reliable energy, efficient
cooling, and strong connectivity gain a durable advantage. They can
deploy and operate AI systems at scale and adapt over time. This leads
to an uneven global landscape, where some regions can expand
continuously while others face hard limits.

Table~\ref{tab:sustainability_constraints} summarizes these
constraints, showing how each one operates, how it limits deployment,
and what mitigation strategies are available. Together, they reinforce
a central point: sustainability is not an optimization goal, but a
feasibility condition.

\begin{table}[t]
\centering
\caption{Sustainability constraints as deployment boundaries for
  AI-oriented data centers. Each constraint operates on a distinct
  timescale and requires different mitigation strategies.}
\label{tab:sustainability_constraints}
\begin{tabular}{p{1.8cm} p{2.2cm} p{3.9cm} p{2.5cm} p{3.4cm}}
\hline
Constraint & Metric & Mechanism & Timescale & Mitigation \\
\hline
Energy
  & MW, grid access
  & Limits scale; grid delays precede deployment
  & Years (grid)
  & On-site generation; demand response \\
Carbon
  & gCO$_2$eq/kWh
  & Determines site eligibility; regulatory exposure
  & Sub-hourly to annual
  & Renewable PPA; carbon-aware scheduling \\
Water
  & L/MWh
  & Limits cooling feasibility; seasonal constraints
  & Seasonal to decadal
  & Liquid cooling; water recycling \\
\hline
\end{tabular}
\end{table}

Ultimately, understanding these constraints is essential for assessing
sovereign AI capacity. What matters is not just what can be designed in
software, but what can actually be built and sustained in the physical
world. The reach of these systems, and therefore the practical scope of
sovereignty, is then determined by the optical networks that connect
them, which are examined in the next section.

\section{Optical Networks as the Backbone of AI Sovereignty}
\label{sec:optical}

Optical networks are what tie AI data centers together into a
functional system. While compute and storage stay anchored to
specific sites, optical transport determines how those resources are
connected, coordinated, and scaled across distance. In practice,
optical infrastructure defines a sovereignty boundary of its own,
separate from compute and energy.

Three factors explain why. First, the physics of propagation sets a
hard lower bound on latency that cannot be reduced by better
hardware or smarter routing~\cite{Singla2014}. Second, network
capacity is limited by spectrum, transceivers, and power, which
ultimately caps how much AI traffic can move between sites. Third,
the physical paths that data follows determine jurisdictional
exposure: which legal systems apply, who can monitor traffic, and
who controls recovery when failures occur. Taken together, these
factors mean that sovereignty depends not only on where data centers
are located, but on how they are connected and governed
\cite{korde2025sovereign}.

\subsection{Latency, capacity, and the physics of optical sovereignty}

The speed of light in fiber, about
$2 \times 10^8$\,m/s, sets a propagation delay of roughly
5\,ms per 1,000\,km~\cite{Singla2014}. This is a hard limit. It does
not improve with faster processors or better protocols. For AI
workloads that depend on tight synchronization across accelerators,
this latency directly defines how far apart those systems can be.

For example, if a training system can tolerate only 1\,ms of
communication delay, then participating nodes must be located within
roughly 100\,km of each other. Beyond that distance, the delay alone
breaks synchronization, regardless of available bandwidth.

Capacity introduces another constraint. As AI systems scale, network
requirements grow from petabit to exabit levels~\cite{Shalf2020,
Chetty2025}. At that point, optical capacity, transceiver density,
and energy efficiency become limiting factors. If the network cannot
move enough data, compute resources sit idle, no matter how powerful
they are~\cite{Tithi2025}. Limits in spectrum, port density, and
power consumption all contribute to these ceilings.

Failures add a third dimension. Fiber cuts, equipment faults, and
regional outages can isolate entire parts of the system~\cite{Gu2020Survey}.
How far those disruptions spread, and how quickly recovery happens,
depends on the structure of the network. A system may meet every
formal requirement for sovereignty and still be fragile if it depends
on a small number of critical links or externally controlled routes
\cite{korde2025sovereign,Cruzes2026}.

\subsection{The scale-up and scale-out boundary}

A key distinction in AI infrastructure is between scale-up and
scale-out networks~\cite{Chetty2025,IEEE_Spectrum2026}. Scale-out
networks connect distributed systems and can tolerate higher
latencies. They support workloads such as inference, data movement,
and replication across regions.

Scale-up networks operate under very different conditions. They
connect accelerators within a training cluster and must support
latencies on the order of microseconds to a few milliseconds
\cite{Naddod2026}. At this scale, propagation delay dominates. A
1\,ms round-trip budget limits the cluster to a radius of about
50\,km.

This makes frontier AI training inherently site-bound. It cannot be
distributed across distant data centers, no matter how much
bandwidth exists between them. Instead, it requires tightly
integrated environments where compute, power, cooling, and
ultra-low-latency optical fabrics are co-located~\cite{Chetty2025}.

The sovereignty implication is straightforward. A region that cannot
build and operate such a scale-up environment cannot host
frontier-level training, regardless of its access to hardware or
energy. Industry analyses show that without improvements in optical
interconnect efficiency, systems remain constrained by short-reach
technologies, which in turn drive extreme rack-level power densities
and complicate cooling~\cite{IEEE_Spectrum2026,Point22026}. This
creates a concentration effect, where only a few locations can meet
all requirements at once.

In practical terms, the physics of latency forces training clusters
into tightly bounded geographic areas. If those conditions cannot be
met locally, sovereignty over frontier training is out of reach,
even if legal and regulatory conditions are satisfied.

\subsection{Network layers and their sovereignty roles}

\begin{figure}[t]
\centering
\includegraphics[width=\linewidth]{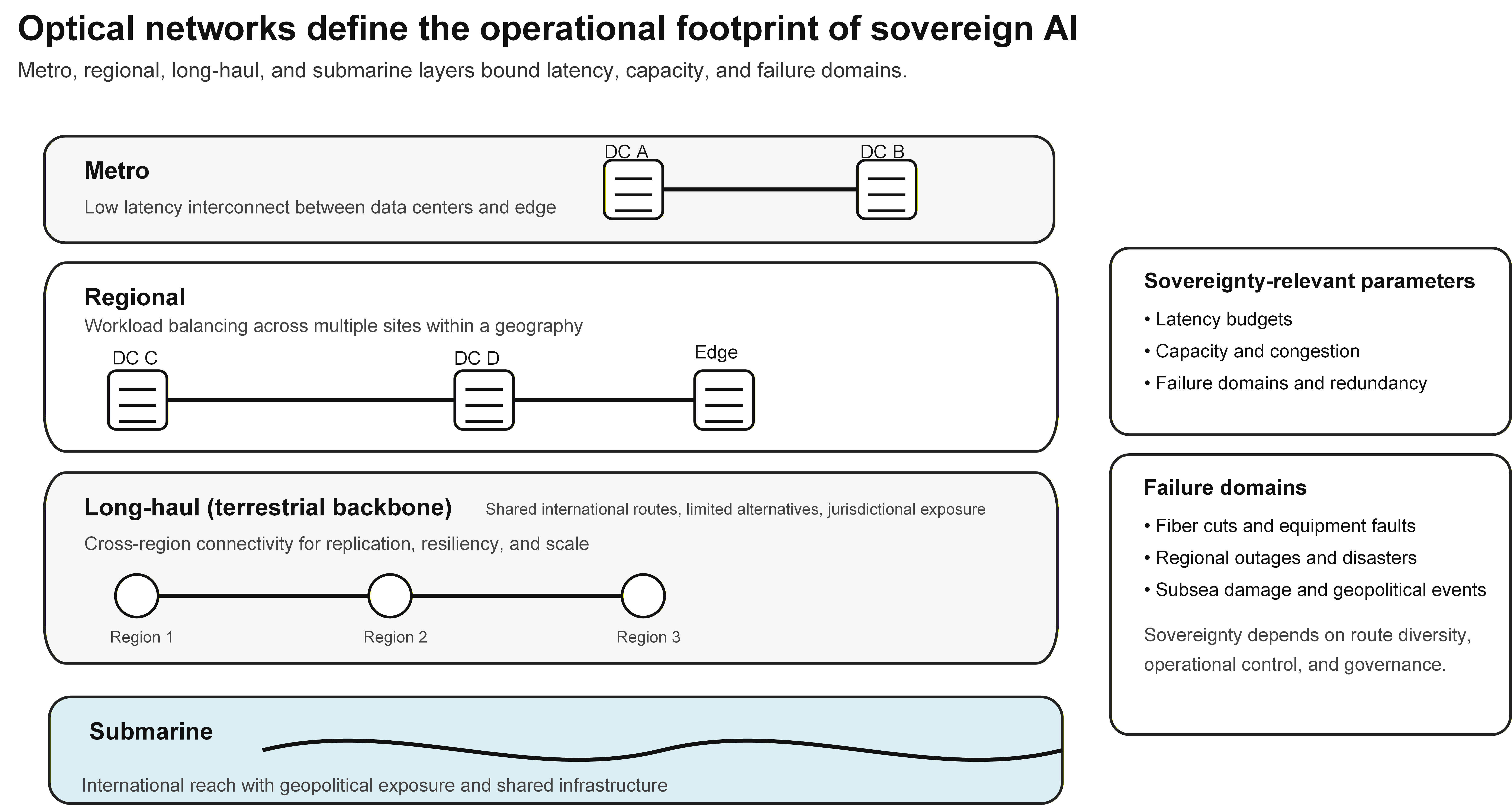}
\caption{Optical networks as the backbone of AI sovereignty. Metro,
  regional, long-haul, and submarine layers each impose distinct
  latency, capacity, and governance constraints that together define
  the operational footprint of distributed AI infrastructure.}
\label{fig:optical_layers}
\end{figure}

Figure~\ref{fig:optical_layers} shows how different layers of the
optical network shape the operational footprint of AI systems.

At the \textbf{metro level}, networks provide low-latency
connectivity within cities, supporting distributed inference and
real-time applications. This layer is typically the most controllable
from a sovereignty perspective, as it is often owned or regulated
domestically.

At the \textbf{regional level}, networks connect multiple sites,
enabling redundancy and workload shifting. This is where
sustainability strategies come into play, allowing workloads to move
between locations with different energy or carbon profiles.

At the \textbf{long-haul level}, networks connect regions and
countries. Latency increases significantly, making these links
unsuitable for real-time applications but still useful for bulk data
movement and replication. Here, sovereignty concerns shift toward
jurisdiction and control, as traffic often crosses multiple legal
domains.

\textbf{Submarine cables} represent the most concentrated point of
risk. They carry large volumes of traffic across limited physical
routes and are difficult to repair. Disruptions can last weeks, and
control over landing points can directly affect access. For regions
dependent on these links, this creates a structural vulnerability
that cannot be mitigated through compute or software alone.

\subsection{Jurisdictional control and the role of operators}

Optical networks also play a direct role in enforcing data
sovereignty. The physical route taken by data determines which legal
frameworks apply while it is in transit. Even if compute and storage
remain within national borders, routing traffic through external
networks can expose it to foreign regulation or interception
\cite{AddOn2024FiberSovereignty}.

This highlights the difference between legal and operational
sovereignty. Legal sovereignty focuses on compliance and data
location. Operational sovereignty depends on control over how data
actually moves. Optical routing is central to this distinction.

Because of this, routing policies and interconnection decisions are
not just technical details. They are sovereignty tools. Regions that
control their own infrastructure and routing policies retain more
control over data in transit than those that depend on external
networks~\cite{GhezChatain2024}.

Telecommunications operators are key in this context. They already
manage the infrastructure that determines routing, interconnection,
and policy enforcement. This places them in a central role in
sovereign AI systems, not just as service providers, but as
operators of critical control points~\cite{ComSoc2025SovereignAI}.

\subsection{Optical reach as a sovereignty multiplier}

Optical reach directly influences how effectively AI resources can
be used. Dense and well-managed networks allow operators to share
capacity across sites, adapt to changing conditions, and maintain
resilience~\cite{AIBottleneck2026,Cruzes2024}. Sparse or fragmented
networks limit flexibility and increase vulnerability to local
failures.

This relationship is not linear. A region with moderate compute but
strong connectivity can be more effective than one with more compute
but weaker networks, because it can coordinate and adapt its
resources more efficiently.

\begin{figure}[t]
  \centering
  \includegraphics[width=\linewidth]{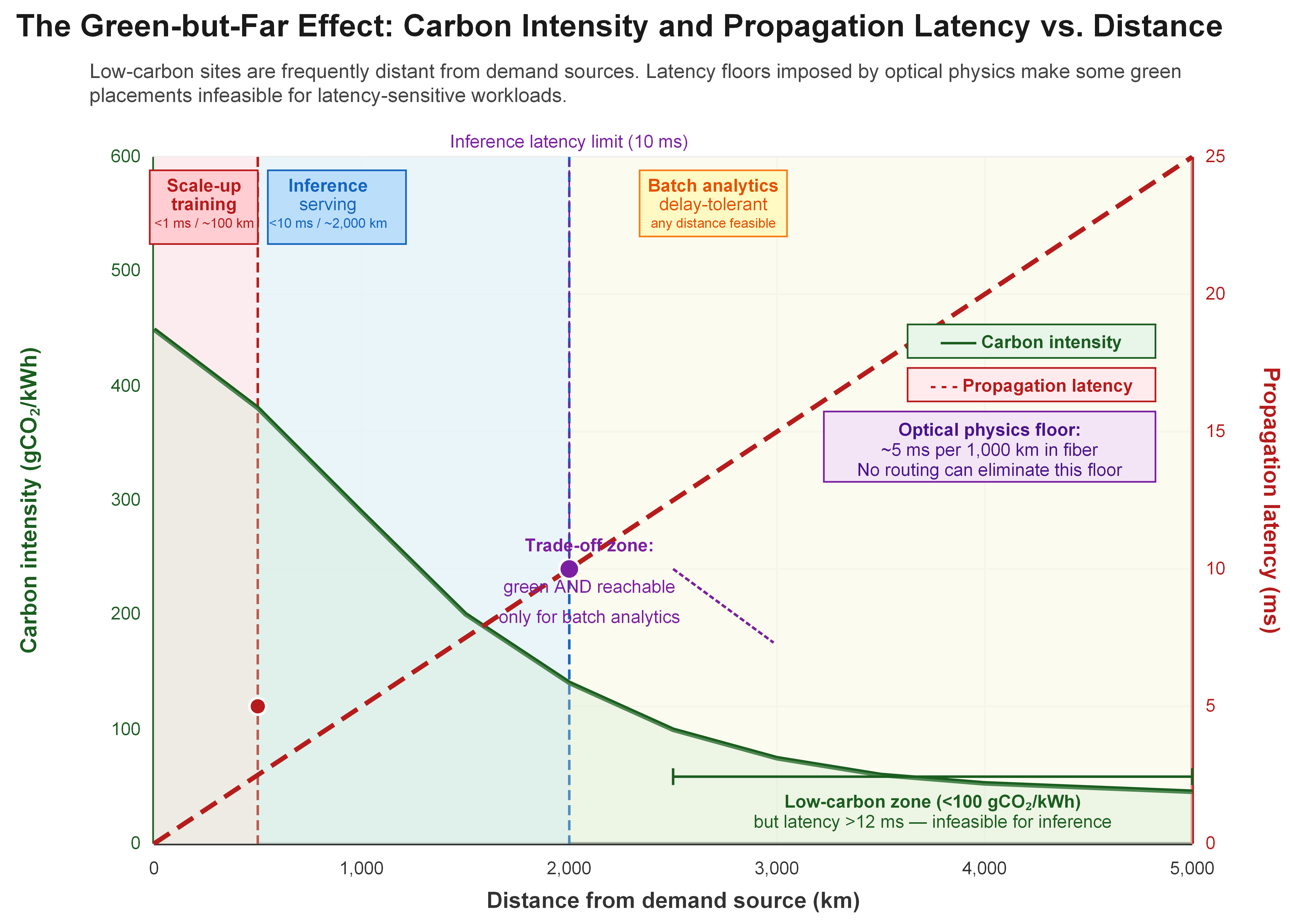}
  \caption{The green-but-far effect: carbon intensity and latency as
  competing placement constraints across AI workloads.}
  \label{fig:green_but_far}
\end{figure}

Figure~\ref{fig:green_but_far} illustrates a key tradeoff between
carbon intensity and latency. Cleaner energy is often located far
from demand, while latency constraints limit how far certain
workloads can be placed. The result is that different workloads have
different degrees of flexibility.

Training workloads, which require very low latency, must remain
close to compute and cannot take advantage of distant low-carbon
sites~\cite{Chetty2025,Tithi2025}. Inference workloads have more
flexibility but are still constrained by latency. Only batch
workloads can be freely moved to optimize for energy or
sustainability~\cite{Hoxha2025LLM}.

This means that sustainability-aware placement is not uniform across
workloads. It is limited by physics and workload requirements, and
must be treated as a structured constraint rather than a general
optimization strategy.

\section{Telemetry as the Foundation of Control and Visibility}
\label{sec:telemetry}

As AI infrastructure grows in scale and coupling, effective control
increasingly depends on continuous and accurate visibility across
compute, network, and energy layers. Telemetry provides this
foundation by exposing the real-time state of data centers, optical
networks, and supporting energy systems, transforming physical
infrastructure from an opaque dependency into an observable and
manageable system~\cite{Cruzes2026,Clemm2015RFC7575}. In the
absence of telemetry, operators are forced to rely on static models,
coarse assumptions, or delayed alarms --- approaches that are
ill-suited to environments characterized by rapid load variation,
tight interdependencies, and narrow operating margins. This section
argues that telemetry is not merely an operational convenience but
a structural prerequisite for sovereign AI control: without the
ability to observe infrastructure state in real time, the ability
to enforce policies, respect sustainability limits, and retain
operational autonomy is nominal rather than effective.

\subsection{What telemetry measures and why cross-layer fusion
matters}

Telemetry in AI infrastructure spans four physically distinct
domains, each with its own instrumentation ecosystem, update rate,
and schema conventions. In AI-oriented data centers, compute and
power telemetry includes rack- and cluster-level power draw,
accelerator utilization, thermal gradients across chip and coolant
boundaries, and UPS state~\cite{Shehabi2024united}. Cooling
telemetry captures coolant supply and return temperatures, flow
rates, make-up water consumption, and ambient wet-bulb temperature,
which together determine evaporative cooling efficiency and water
permit exposure. In optical networks, telemetry captures signal
quality indicators including optical signal-to-noise ratio and
pre-FEC bit error rate, spectrum occupancy, per-link latency, and
fault and alarm state~\cite{Cruzes2026}. Grid sustainability
telemetry adds marginal carbon intensity, energy headroom, and
on-site generation and storage state, sourced from grid signal
providers and building energy management systems.

Collecting these streams in isolation provides limited value. The
operational significance of telemetry emerges from cross-layer
fusion: combining measurements across domains to reveal causal
chains that remain invisible within siloed monitoring
systems~\cite{Gu2020Survey}. Consider three examples. First, a
thermal event at a specific rack --- detected through coolant
return temperature rising above a threshold --- may require
workload migration to a cooler site, which increases inter-site
traffic, which in turn loads optical links that were near
congestion. A monitoring system that observes only compute or
only network state cannot anticipate this cascade; a cross-layer
system can detect the thermal precursor and pre-emptively reserve
routing capacity before the migration is triggered. Second, a
spike in grid carbon intensity at one site, if detected alongside
available power headroom at a low-carbon alternative site and
sufficient optical capacity between them, creates an actionable
window for sustainability-driven workload shifting that no
single-domain monitoring system can identify. Third, a fiber cut
that reduces available optical capacity between two sites may
force workloads to remain at high-carbon or water-stressed sites
that would otherwise have been vacated, because the alternative
routing path no longer exists. Cross-layer telemetry makes this
constraint visible before a placement decision is made on
incorrect assumptions about network availability.

\subsection{Streaming architectures and the standardization
challenge}

Making telemetry operationally useful requires streaming
architectures that push measurements continuously into monitoring
and control platforms rather than relying on periodic polling or
coarse-grained summaries~\cite{Cruzes2026,TMF398}. Polling
intervals of minutes or hours are incompatible with AI environments
where power draw and thermal conditions change on sub-second
timescales during collective communication
operations~\cite{choukse2025powerstab}. Streaming telemetry
enables near-real-time observability in which changes in load,
failures, or environmental conditions are detected within the
timescales at which corrective actions remain effective, allowing
automation systems and agents to adjust configurations before
local degradations escalate into system-wide disruptions.

The standardization challenge underlying this shift is significant
and constitutes a practical sovereignty concern in its own right.
Each physical domain exposes telemetry through a different protocol
stack, schema convention, and transport mechanism, reflecting the
fact that these domains evolved independently under separate
engineering communities and standards bodies. Integrating them
into a unified observability plane therefore requires explicit
normalization work that cannot be delegated to any single vendor
or platform.

Optical networks are the most mature domain from a telemetry
standardization perspective. They increasingly support OpenConfig
YANG models --- a vendor-neutral, open-source data modeling
framework that defines device configuration and state in a
structured, machine-readable schema~\cite{openconfig2016} ---
streamed over gNMI (gRPC Network Management Interface), a
high-rate, bidirectional telemetry transport built on
gRPC~\cite{Shakir2018gNMI}. gRPC itself operates over HTTP/2 and
uses Protocol Buffers for compact binary encoding, enabling
sub-second streaming of optical KPIs such as signal-to-noise ratio
(OSNR), bit error rate (BER), and per-channel power from
reconfigurable optical add-drop multiplexers (ROADMs),
transponders, and amplifiers directly into control
platforms~\cite{Cruzes2026}. Despite this progress, vendor
implementations vary significantly in schema coverage and update
rate fidelity, and many deployed devices still expose only
NETCONF~\cite{rfc6241} or Simple Network Management Protocol
(SNMP)~\cite{RFC1157} interfaces that require polling rather than
streaming.

Compute and power domains use a different and less unified set of
interfaces. IPMI (Intelligent Platform Management Interface) is a
hardware-level specification that allows out-of-band monitoring
and management of server hardware --- including power draw, thermal
state, and hardware status --- independently of the operating
system~\cite{IPMI2013}. Redfish~\cite{DMTF2024Redfish} is a more
modern REST-based standard developed by the DMTF (Distributed
Management Task Force) that replaces IPMI for server management,
providing structured JSON-based access to power consumption,
thermal margins, and hardware inventory through a well-defined
API. BMC APIs (Baseboard Management Controller APIs) are
vendor-specific interfaces exposed by the embedded management
processor present on most server motherboards; they provide
low-level hardware telemetry but differ significantly across
manufacturers in schema, update rate, and supported metrics,
complicating cross-vendor normalization~\cite{DMTF2024Redfish}.
In practice, Redfish is preferred where available because it
provides freshness metadata and structured schemas compatible with
automated ingestion; IPMI and proprietary BMC APIs require
additional normalization before their outputs can be admitted to
a sovereignty-aware control system.

Cooling and building management systems operate under a further
distinct set of protocols designed for facilities engineering
rather than real-time infrastructure control. BACnet (Building
Automation and Control Networks, standardized as ASHRAE 135 and
ISO 16484-5) is the dominant protocol for building automation
systems, covering HVAC, cooling, lighting, and power distribution;
it supports both IP-based (BACnet/IP) and legacy serial
(BACnet MS/TP) transport, with read and write operations organized
around a structured object model~\cite{ASHRAE135}. Modbus is an
older serial communication protocol widely used in industrial
control and facilities management for reading sensor data from
cooling equipment, power distribution units, and environmental
monitors; it is simple and robust but lacks structured schemas,
requiring manual mapping of register addresses to physical
measurements~\cite{Modbus2012}. Neither BACnet nor Modbus was
designed for the sub-second update rates or the time-aligned,
freshness-certified streams that real-time AI infrastructure
control requires, and both need adaptation layers before their
outputs can be fused with optical and compute telemetry.

Grid sustainability signals introduce a fourth protocol ecosystem
governed by commercial APIs rather than open engineering
standards. WattTime~\cite{Cote2025} is a service that provides
real-time and forecast marginal carbon intensity signals --- the
emissions rate of the next unit of electricity drawn from the
grid at a given location and time --- sourced from grid operator
dispatch data; it is used to determine whether a workload
placement decision would consume low-carbon or high-carbon
electricity at execution time. Electricity Maps provides similar
signals with broader geographic coverage and both real-time and
historical data access. Both services expose REST APIs with their
own authentication schemes, rate limits, schema conventions, and
update intervals (typically five minutes for real-time signals),
and neither was designed with the freshness certification or
provenance tracking requirements of a sovereign control system in
mind. Grid operator APIs, where available, can provide more
authoritative signals but vary significantly by jurisdiction in
availability, access terms, and data
format~\cite{Cote2025,carbonawaresdk}.

The result is a heterogeneous telemetry landscape in which four
physically distinct domains --- optical networking, compute and
power, cooling and facilities, and grid sustainability --- each
speaks a different protocol, operates on a different update
cadence, and exposes data in a different schema. Integrating
these streams into the unified state representation
$\boldsymbol{\theta}(t)$ required by the agentic control system
is not a commodity integration task but a sovereignty-relevant
engineering challenge: an operator who delegates this integration
layer to a vendor-supplied management platform or a cloud-based
observability service has effectively delegated part of their
situational awareness, and therefore part of their operational
sovereignty, to an external actor whose behavior and data policies
they do not fully govern. The ingestion layer that performs this
integration --- schema normalization, timestamp alignment,
freshness validation, and confidence scoring --- is examined in
architectural detail in Section~\ref{sec:reference_architecture}.

\subsection{Data quality, freshness, and AI-specific failure modes}

The effectiveness of telemetry depends critically on data quality,
freshness, and synchronization, and the failure modes in AI
environments are more consequential than in conventional data
center operations~\cite{Cruzes2026}. Three failure modes deserve
explicit treatment.

\textbf{Stale carbon intensity signals} are among the most
operationally significant. Grid carbon intensity can change
substantially within a single optimization interval --- shifting
from below to above a regulatory threshold, or vice versa, within
minutes during generation dispatch transitions. A placement
decision made on a carbon intensity reading that is 15 minutes
old may allocate a workload to a site whose grid has since
exceeded the operator's carbon policy limit. Because the
hard-constraint model eliminates non-compliant configurations
entirely rather than penalizing them gradually, this staleness
does not produce a slightly suboptimal outcome: it produces a
constraint violation that may have regulatory consequences.
Freshness thresholds for carbon intensity signals must therefore
be set relative to the rate of change of grid dispatch, not
relative to the convenience of the data provider's update
schedule.

\textbf{Misaligned timestamps across domains} corrupt cross-layer
causal inference. Power meters may update every 30 seconds,
carbon intensity signals every 5 minutes, water consumption
readings every hour, and optical performance monitoring every few
seconds. If these streams are fused without explicit timestamp
alignment, a power spike that occurred at 14:03 may be correlated
with a carbon intensity reading from 13:58 and a water consumption
reading from 14:00, producing a misleading picture of the system
state at any given moment. In AI training environments where
collective communication produces power spikes lasting
milliseconds to seconds~\cite{choukse2025powerstab}, misalignment
of even a few seconds can cause the monitoring system to associate
a thermal event with the wrong workload phase, leading to
incorrect fault attribution and misdirected corrective action.

\textbf{Missing or degraded telemetry from critical paths}
requires a disciplined degradation policy rather than silent
substitution with last-known-good values. When alarm state on an
optical link is unavailable because a management plane fault has
interrupted the telemetry stream, silently assuming the link is
healthy can cause the placement optimizer to route traffic over a
link that is in active protection switching, producing an actual
outage from an attempted optimization. A principled degradation
policy distinguishes between parameters for which conservative
substitution is safe --- power headroom can be replaced by a
lower bound, slowly varying water consumption by the most recent
reading with a widened uncertainty interval --- and parameters
for which no safe substitute exists and optimization should be
held until fresh telemetry is restored.

\subsection{Telemetry as a sovereignty boundary}

As AI infrastructure spans multiple operators, vendors, and
jurisdictions, telemetry itself becomes a sovereignty boundary
whose governance requires as much attention as the physical
infrastructure it monitors. Automation and agentic control across
domain boundaries require sharing fine-grained telemetry ---
power state, network topology, workload placement, carbon
intensity --- with counterparts in adjacent domains. Unrestricted
sharing of this data, however, exposes sensitive operational
information: traffic matrices reveal workload patterns and
customer behavior; power draw profiles reveal training schedules
and model sizes; fault histories reveal infrastructure
vulnerabilities. In environments where AI infrastructure is
operated by or on behalf of government agencies, critical
services, or strategically sensitive industries, this exposure is
a direct sovereignty risk.

Recent work on governed telemetry data sharing shows that
cross-domain automation and sovereignty can be reconciled through
policy-enforced access controls rather than through
isolation~\cite{Mitrovska2025DataGovernance}. By attaching usage
policies directly to telemetry streams --- specifying which
stakeholders can access which measurements, under what time
constraints, at what geographic granularity, and for what
automated purposes --- it becomes possible to support cross-domain
control while retaining operational authority and regulatory
compliance. A telemetry stream that exposes aggregate carbon
intensity at a regional level for sustainability scheduling,
while withholding site-level power draw and workload composition
from external systems, enables cooperation without disclosure.
Enforcing these policies at the stream level rather than through
post-hoc access controls ensures that sovereignty constraints are
embedded in the infrastructure workflow rather than dependent on
contractual agreements between operators.

This approach has a practical implication for architecture:
telemetry governance must be treated as a first-class design
concern, not an afterthought applied to an already-designed
monitoring system. An organization that designs its telemetry
pipeline without sovereignty constraints and then attempts to
add them later will find that the pipeline's schema, transport
protocols, and aggregation logic have already embedded
assumptions about data sharing that are difficult to retrofit.
Sovereignty-preserving telemetry requires that access policy,
data minimization, and jurisdictional routing be specified at
the same time as measurement schema and update rate.

To state this plainly for non-specialist readers: telemetry is
not merely technical data about infrastructure operation --- it
is also information about how and when systems are used, where
workloads run, and how resources are consumed. If shared without
governance, it reveals sensitive operational details. Treating
access policy, data minimization, and jurisdictional routing as
first-class design requirements --- specified alongside
measurement schema and update rate --- is therefore a sovereignty
obligation, not an afterthought.

\subsection{Telemetry as the foundation of operational sovereignty}

In the framework developed in this tutorial, telemetry bridges
the gap between legal sovereignty --- which concerns ownership,
jurisdiction, and formal control rights --- and operational
sovereignty, which requires the ability to observe infrastructure
state, make decisions based on local conditions, and execute
actions within locally defined policies. Legal sovereignty without
telemetry is nominal: an operator may formally control AI
infrastructure while lacking the real-time visibility needed to
enforce carbon limits, detect constraint violations, or respond
to failures before they propagate. Operational sovereignty
requires, at minimum, that the operator can see what the
infrastructure is doing at the timescale at which it changes.

Telemetry also strengthens sovereignty in heterogeneous
environments where full asset ownership is not achievable. A
multi-vendor AI data center that exposes real-time power, cooling,
and network telemetry into a unified control plane can enforce
local policies regardless of equipment origin. A facility that
relies on foreign-manufactured accelerators but monitors their
power draw, thermal state, and utilization locally retains
operational authority over how those accelerators are used, even
when it does not control their design or supply chain. This shifts
the locus of sovereignty from asset ownership --- an increasingly
unrealistic standard given global supply chains --- to operational
capability: the ability to observe, decide, and act within locally
defined constraints~\cite{CloudSecurity2025}.

The full realization of this capability requires that telemetry
feeds not only human operators but automated reasoning systems
that can act on it at the timescales AI infrastructure demands.
The agentic AI systems and digital twins that close this loop are
examined in the following section.

\section{Agentic AI and Digital Twins for Sovereign Operation}
\label{sec:agentic}

As AI infrastructure spans data centers, optical networks, and
energy systems, manual operation no longer scales. The number of
interacting variables --- power draw, thermal state, grid carbon
intensity, optical link utilization, workload placement, water
permit exposure --- combined with tight timing, safety, and
sustainability constraints, exceeds what human operators can manage
reliably in real time~\cite{Clemm2015RFC7575}. Agentic AI addresses
this challenge by structuring control logic into autonomous but
bounded agents, each responsible for observing system conditions,
reasoning about feasible actions, and proposing decisions within
explicitly defined limits~\cite{Cruzes2026}. An \emph{agent} in
this context is a software component that perceives its environment
through telemetry, evaluates options against a policy specification,
and proposes actions --- but does not execute them unilaterally.

Critically, agentic AI does not replace human authority or policy
decision making: it structures the \emph{execution} of decisions
under predefined policies, safety constraints, and operational
boundaries, making automation auditable and its scope of authority
explicit. This section argues that agentic AI and digital twins,
grounded in the telemetry infrastructure described in
Section~\ref{sec:telemetry}, are the operational mechanisms through
which legal sovereignty is converted into genuine operational
sovereignty.

\subsection{Agent architecture and coordination}

\begin{figure*}[t]
  \centering
  \includegraphics[width=\linewidth]{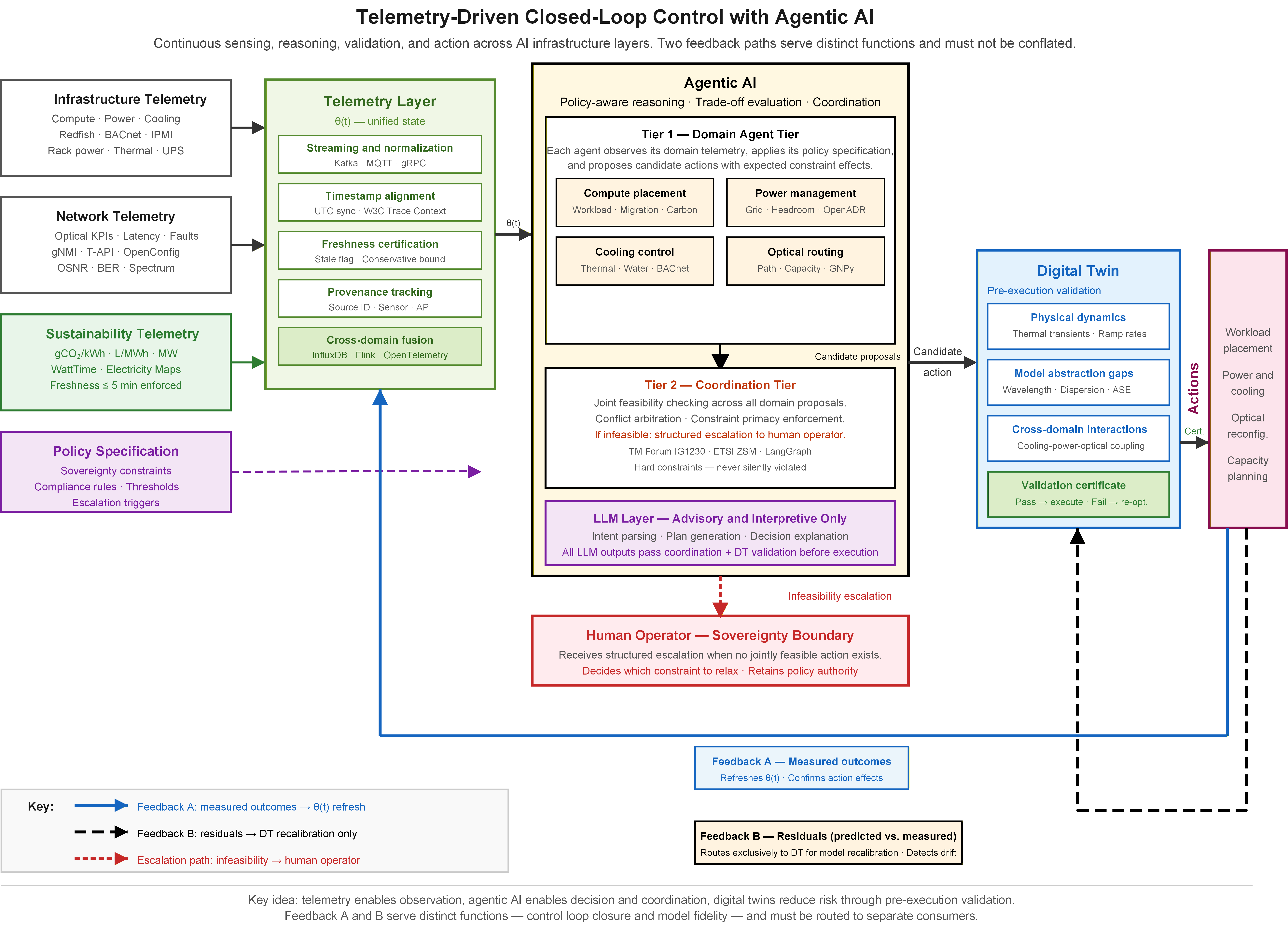}
  \caption{%
    \textbf{Telemetry-driven closed-loop control with agentic AI
    for sovereign infrastructure operation.}
    Three telemetry streams --- infrastructure, network, and
    sustainability --- feed the observability layer, which
    assembles the unified state representation
    $\boldsymbol{\theta}(t)$ through streaming normalization,
    timestamp alignment, freshness certification, and
    cross-domain fusion.
    A locally defined policy specification bounds agent authority
    throughout.
    The agentic AI block organizes reasoning into two tiers:
    a domain agent tier (compute placement, power management,
    cooling control, optical routing) that proposes candidate
    actions within bounded scopes, and a coordination tier that
    evaluates joint feasibility across all proposals.
    When no jointly feasible action exists, the coordination tier
    escalates to the human operator (red dashed path) with a
    structured description of the binding constraint conflict ---
    a sovereignty mechanism that prevents silent policy violation.
    The LLM layer operates in an advisory and interpretive role
    only; all outputs are validated before execution.
    The digital twin validates candidate actions against physical
    dynamics, model abstraction gaps, and cross-domain
    interactions before issuing a validation certificate.
    Only certificate-bearing actions reach southbound interfaces.
    \textbf{Feedback~A} (solid blue) returns measured outcomes
    to the telemetry layer to refresh $\boldsymbol{\theta}(t)$.
    \textbf{Feedback~B} (dashed orange) returns prediction
    residuals exclusively to the digital twin for model
    recalibration and drift detection.
    These two paths serve distinct functions --- control loop
    closure and model fidelity --- and must not be conflated.%
  }
  \label{fig:telemetry_agentic_loop}
\end{figure*}

Agent-based decision making enables coordination across
infrastructure layers without centralizing all control logic in a
single controller~\cite{Cruzes2026,TMF398}.
Figure~\ref{fig:telemetry_agentic_loop} illustrates the full
closed-loop architecture developed in this section.

To make this flow explicit, the architecture can be read from left
to right as a sequence of transformations from observation to
validated action. Three distinct telemetry sources form the input
layer. \textbf{Infrastructure telemetry} captures compute, power,
and cooling state, including rack power draw, thermal conditions,
and hardware status through interfaces such as Redfish and
IPMI~\cite{DMTF2024Redfish}. \textbf{Network telemetry} captures
optical-layer and transport metrics, including latency, faults, and
physical-layer indicators such as OSNR, BER, and spectrum
utilization via streaming interfaces such as gNMI and OpenConfig
models~\cite{Shakir2018gNMI}. \textbf{Sustainability telemetry}
captures carbon intensity, water utilization, and energy headroom,
typically sourced from external grid interfaces and required to
meet strict freshness bounds for decision
making~\cite{Cote2025,carbonawaresdk}.

These streams are ingested into the \textbf{telemetry layer}, which
constructs the unified system state $\boldsymbol{\theta}(t)$
through a series of processing steps: streaming and normalization
of heterogeneous inputs, timestamp alignment to a common time base,
freshness certification to bound decision risk, provenance tracking
to preserve source traceability, and cross-domain fusion to produce
a consistent multi-layer representation of infrastructure state.
These mechanisms are typically implemented using streaming and
observability frameworks such as Kafka and
OpenTelemetry~\cite{raptis2023kafka,otel2023}.

The unified state feeds the \textbf{agentic AI layer}, organized
into two tiers. In the \textbf{Tier~1 domain agent layer},
specialized agents perform bounded reasoning within their
respective domains: compute placement (workload allocation and
migration), power management (grid interaction and headroom
control), cooling control (thermal and water system actuation),
and optical routing (path selection and capacity management). Each
agent proposes candidate actions together with an explicit account
of the constraints it satisfies and the trade-offs it introduces.

These proposals are evaluated by the \textbf{Tier~2 coordination
layer}, which performs joint feasibility checking across domains
using policy frameworks such as TM Forum IG1230 and ETSI
ZSM~\cite{TMF398} and graph-based agent orchestration frameworks.
Conflicting proposals are not silently resolved: when no feasible
joint action exists, the system produces a structured infeasibility
report and escalates to the human operator.

The \textbf{LLM layer} operates strictly in an advisory and
interpretive role, translating human intent into structured
objectives and explaining decisions, consistent with prior work on
LLM-assisted network automation~\cite{Cruzes2025llm}. All
LLM-generated outputs are subject to coordination-layer constraint
enforcement and digital twin validation before execution.

Candidate actions are then evaluated by the \textbf{digital twin},
which performs pre-execution validation across physical dynamics,
model abstraction gaps, and cross-domain interactions, as commonly
adopted in digital twin frameworks for infrastructure
systems~\cite{Cruzes2026}. Only actions that pass this validation
stage receive a \emph{validation certificate} and are allowed to
reach execution interfaces.

When no valid action can be constructed, control passes to the
\textbf{human operator}, who defines the sovereignty boundary of
the system. The operator receives a structured description of
constraint conflicts and retains authority to relax or prioritize
constraints, ensuring that policy decisions remain explicitly
human-controlled.

Two feedback paths close the loop. \textbf{Feedback~A} carries
measured outcomes back to the telemetry layer to update
$\boldsymbol{\theta}(t)$ and confirm action effects.
\textbf{Feedback~B} carries residuals between predicted and
observed outcomes exclusively to the digital twin, enabling model
recalibration and drift detection. These two feedback paths serve
distinct functions and must remain strictly separated.

\subsection{Digital twins: pre-execution validation and model
fidelity}

Digital twins play a critical role in validating agent decisions
before they affect live infrastructure~\cite{Cruzes2026}. A digital
twin maintains a synchronized computational model of the physical
infrastructure --- compute sites, power delivery systems, cooling
circuits, and optical network topology --- and uses it to simulate
the expected consequences of a proposed action before that action
is executed. This pre-execution validation is essential in tightly
coupled environments where missteps propagate rapidly: a workload
migration that overloads a cooling circuit can trigger thermal
throttling across an entire rack; a routing change that exceeds
optical link capacity can degrade signal quality below the FEC
threshold and cause a link failure; a power allocation change that
violates a demand response contract can impose financial penalties
and trigger grid-side protection actions.

The digital twin operates on three classes of constraints that the
optimization model alone cannot fully enforce. The optimization
layer is formulated as a mixed-integer linear program (MILP), which
represents the system using linear constraints and discrete
decision variables to enable tractable, global
optimization~\cite{Bixby2012}. While this formulation is effective
for capturing steady-state feasibility and coordinating decisions
across domains, it necessarily abstracts away non-linear dynamics,
transient effects, and higher-order physical interactions.

First, \textbf{physical dynamics} that are abstracted away in the
MILP formulation: thermal transients, power ramp rates, coolant
flow stabilization times, and optical amplifier gain settling times
all evolve on timescales shorter than the optimization cycle and
can cause constraint violations during the transition to a new
configuration even when the steady-state target is compliant.
Second, \textbf{model abstraction gaps}: the optimization graph
represents the optical network as a set of logical edges with
capacity and delay parameters, but the physical layer has
wavelength assignments, amplifier cascades, and chromatic
dispersion margins that determine whether a proposed routing is
physically realizable at the optical layer. Third,
\textbf{cross-domain interactions} that are too complex to
enumerate as MILP constraints: the interaction between a specific
cooling configuration and the ambient temperature forecast at a
given site, for example, may determine whether a proposed power
increase is thermally safe, but encoding this interaction as a
linear constraint requires approximations that the digital twin
can evaluate more faithfully using a higher-fidelity thermal model.

Maintaining the fidelity of the digital twin over time is as
important as its initial accuracy. Physical infrastructure changes
--- equipment replacements, cooling system upgrades, fiber route
additions, grid connection modifications --- must be reflected in
the twin promptly, or the twin will validate actions against a
model of infrastructure that no longer exists. Model drift, in
which the twin's representation progressively diverges from
physical reality, is a failure mode with direct sovereignty
consequences: an operator who believes their digital twin provides
accurate pre-execution validation but whose twin has drifted is
making decisions on the basis of an inaccurate model without
knowing it. Continuous reconciliation between twin predictions and
measured telemetry outcomes, flagging cases where the twin
predicted one outcome and the infrastructure produced another, is
the primary mechanism for detecting and correcting model
drift~\cite{Cruzes2026,Kiasari2026AgenticGrid}.

\subsection{Large language models at the human-automation
interface}

Large language models contribute at the interface between human
intent and automated control, occupying a specific and bounded role
that is distinct from both policy specification and action
execution~\cite{Cruzes2025llm}. Operators typically express
operational goals in abstract terms: reduce the carbon footprint
of the training cluster, maintain inference latency below 50\,ms
during the maintenance window, or prepare the facility for a
predicted grid stress event at 18:00. These expressions are
underspecified from the perspective of the agentic control system:
they do not identify which workloads to migrate, which sites are
eligible, which constraints take precedence when trade-offs arise,
or what the acceptable degradation in secondary objectives is.

Large language models (LLMs) bridge this gap by translating
natural language intent into structured objective specifications
and constraint parameterizations that the agentic system can
evaluate against real-time telemetry and digital twin
models~\cite{Cruzes2026}. They also enhance operational
transparency by generating human-readable explanations of
decisions: why a specific workload was placed at a specific site,
which constraints were binding, what trade-offs were made between
carbon and latency, and what alternative actions were considered
and rejected. This explanatory capability is particularly
important for sovereignty, because it makes automated decisions
auditable: an operator can reconstruct the reasoning behind a
configuration change and verify that it was consistent with
declared policy.

The boundary around LLM involvement must be drawn carefully. LLMs
are probabilistic systems whose outputs are not guaranteed to be
correct, consistent, or policy-compliant. Allowing LLM outputs to
drive infrastructure actions directly --- without validation
through the deterministic constraint-checking of the agentic
system and the pre-execution simulation of the digital twin ---
introduces a failure mode in which plausible-sounding but
incorrect instructions are executed on live
infrastructure~\cite{Cruzes2025llm}. The appropriate role is
therefore advisory and interpretive: LLMs propose structured
objectives and explain decisions, but every proposed action passes
through the agent coordination layer and digital twin validation
before execution. This architecture preserves the intelligibility
benefits of natural language interaction while containing the risk
of LLM error within the safety envelope provided by deterministic
validation.

\subsection{From manual operation to sovereign autonomous control}

The transition from manual operation to agentic control represents
a qualitative change in the nature of operational sovereignty, not
merely an increase in automation level.
Table~\ref{tab:control_approaches} characterizes this transition
across five dimensions: observability, pre-execution validation,
decision autonomy, human oversight requirement, and primary failure
mode. The comparison shows that higher autonomy is not achieved by
reducing human oversight but by grounding decisions in continuous
telemetry, validating them against physical models, and
constraining their scope within explicitly specified policies. The
primary failure mode shifts from human error and reaction-time
limits under manual operation, to rule brittleness and
edge-case blindness under rule-based automation, to model drift
and coordination failure under agentic control --- a progression
that reflects increasing capability alongside shifting risk
profiles that require different governance responses.

\begin{table}[t]
\centering
\small
\caption{Comparison of control approaches for AI infrastructure
  sovereignty. Higher autonomy shifts the primary failure mode
  from human reaction limits to model fidelity and coordination,
  requiring governance responses focused on twin reconciliation
  and escalation path design rather than operator training.}
\label{tab:control_approaches}
\begin{tabular}{p{2.7cm} p{2.0cm} p{2.3cm} p{1.7cm} p{2.2cm}
  p{2.9cm}}
\hline
Approach & Observability & Validation & Autonomy
  & Human oversight & Primary failure mode \\
\hline
Manual operations
  & Low, periodic & None & Low & Continuous
  & Reaction time; fatigue \\
Rule-based automation
  & Medium, polled & Rule-check only & Medium
  & Exception-driven & Rule brittleness; edge cases \\
Agentic with digital twins
  & High, streaming & Pre-execution simulation & High
  & Escalation-driven & Model drift; coordination failure \\
\hline
\end{tabular}
\smallskip

\noindent\footnotesize Entries are qualitative characterizations
of typical deployments within each approach; actual observability,
validation depth, and oversight requirements vary with
implementation scale and policy specification.
\end{table}

\subsection{Agentic control as operational sovereignty}

As defined in Section~\ref{sec:intro}, this work distinguishes
between \emph{legal sovereignty} and \emph{operational
sovereignty}. Legal sovereignty refers to formal control over
ownership, jurisdiction, and access rights, while operational
sovereignty denotes the ability to observe, decide, and act on
infrastructure state in practice. In the framework developed in
this tutorial, agentic AI and digital twins are the mechanisms
that transform legal sovereignty into operational sovereignty.

Legal sovereignty alone does not guarantee the ability to observe
infrastructure state, enforce sustainability constraints, or
respond to failures in real time. An operator may formally own AI
infrastructure and therefore satisfy legal sovereignty, yet lack
operational sovereignty if they depend on external systems or
providers to monitor, validate, or execute control actions. In
such cases, control exists at the level of policy but not at the
level of execution.

Agentic control closes this gap by establishing operational
sovereignty through locally grounded control loops. Decisions are
derived from locally collected telemetry rather than
vendor-supplied state estimates. Candidate actions are validated
against locally maintained digital twin models rather than
external simulation services. Executed actions are constrained by
locally defined policies rather than platform defaults. When
automated competence reaches its limit, the system escalates to
local human authority with structured explanations rather than
relying on external support~\cite{Nalage2025}. This combination
--- local observation, local validation, local policy enforcement,
and structured escalation --- constitutes operational sovereignty
in practice.

This interpretation of operational sovereignty aligns with
policy-aware agentic control frameworks developed in adjacent
domains, including cloud data pipeline governance, intelligent
infrastructure monitoring, and smart grid
control~\cite{Kirubakaran2025,Kiasari2026AgenticGrid}. The
following section integrates the physical infrastructure examined
in Sections~\ref{sec:datacenters} and~\ref{sec:optical}, the
telemetry foundation of Section~\ref{sec:telemetry}, and the
agentic control architecture developed here into a unified
reference architecture for sovereign AI infrastructure.

\section{Reference Architecture for AI Infrastructure Sovereignty}
\label{sec:reference_architecture}

The preceding sections have examined the physical, operational, and
control dimensions of AI infrastructure sovereignty in sequence: data
center sustainability limits, optical network boundaries, telemetry
as the observability foundation, and agentic AI with digital twins as
the control mechanism. This section integrates these elements into a
unified reference architecture that specifies not only what components
are required but how they connect, what flows across their interfaces,
how the system degrades gracefully under partial failure, and how
sovereignty level can be assessed against a structured framework. The
architecture is not prescriptive of specific vendors or
implementations; it is prescriptive of functional responsibilities,
interface contracts, and sovereignty properties that any compliant
implementation must satisfy.

\subsection{Architectural overview and design principles}

\begin{figure*}[t]
  \centering
  \includegraphics[width=\linewidth]{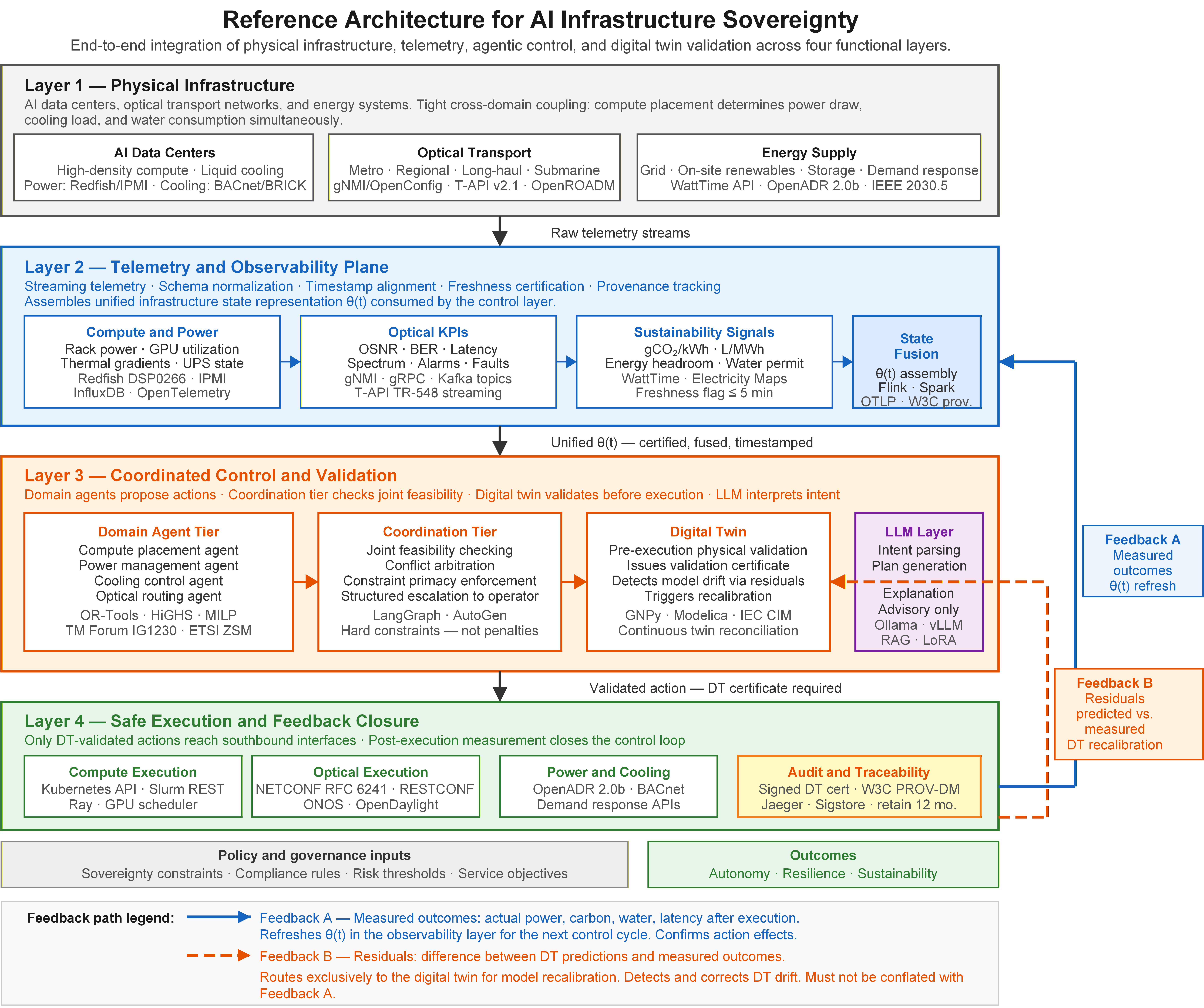}
  \caption{%
    \textbf{Reference architecture for AI infrastructure sovereignty,
    organized across four functional layers.}
    The physical layer (Layer~1) provides AI data centers, optical
    transport, and energy systems with tight cross-domain coupling.
    The observability layer (Layer~2) streams, normalizes, and
    time-aligns cross-domain telemetry into the unified
    infrastructure state representation $\boldsymbol{\theta}(t)$.
    The control layer (Layer~3) applies domain agent reasoning,
    coordination-tier joint feasibility checking, digital twin
    pre-execution validation, and LLM-assisted intent parsing to
    produce safe, policy-compliant candidate actions.
    The execution layer (Layer~4) applies those actions through
    domain-specific southbound interfaces and enforces the
    invariant that only digitally twin-validated actions reach
    live infrastructure.
    Two distinct feedback paths close the loop:
    \textbf{Feedback~A} (solid blue) returns measured outcomes to
    the observability layer to refresh $\boldsymbol{\theta}(t)$;
    \textbf{Feedback~B} (dashed orange) returns prediction
    residuals exclusively to the digital twin for model
    recalibration and drift detection.
    These two streams serve different functions --- control
    versus model fidelity --- and must not be conflated.
    Sovereignty properties are realized at each layer and depend
    on the integrity of the interfaces between them.%
  }
  \label{fig:ref_arch}
\end{figure*}

Figure~\ref{fig:ref_arch} presents the reference architecture as a
closed-loop control system organized across four functional layers.
The layers are vertically integrated through defined interfaces and
horizontally coupled through the shared telemetry plane that makes
the state of each layer visible to all others. A critical
architectural distinction governs the feedback paths that close the
loop from Layer~4 back into the system. Two physically separate
feedback streams serve fundamentally different functions and must not
be conflated. \textbf{Feedback~A} (solid blue) carries measured
outcomes --- actual power draw, carbon consumption, water usage, link
utilization, and workload placement observed after execution --- back
to the observability layer, where they refresh the unified state
representation $\boldsymbol{\theta}(t)$ for the next control cycle.
This is the primary closed-loop signal confirming whether executed
actions produced their intended effects. \textbf{Feedback~B} (dashed
orange) carries residuals --- the difference between digital twin
predictions and measured outcomes --- exclusively to the digital twin
for model recalibration. This stream provides the ground-truth signal
needed to detect and correct model drift before it compromises
validation reliability. Routing Feedback~B into the observability
layer rather than the digital twin, or conflating it with Feedback~A
into a single monitoring stream, destroys the model drift detection
capability that is essential for maintaining twin fidelity over time.
These two streams must be maintained separately and routed to their
respective consumers at all times, including during partial failures
of the telemetry or execution infrastructure.

Three design principles govern the architecture. The first is
\emph{local grounding}: every decision made by the control layer
must be based on telemetry collected locally, validated against
models maintained locally, and executed within policies defined
locally. Dependence on external state estimates, remote simulation
services, or platform-provider control logic at any of these steps
reduces operational sovereignty proportionally. The second is
\emph{constraint primacy}: sustainability and safety constraints are
encoded as hard boundaries that eliminate non-compliant actions from
the feasible set rather than as penalties that can be traded against
performance objectives. This encoding reflects the legal and physical
reality that regulatory maxima are not preferences but absolute
limits whose violation carries consequences that cannot be offset by
gains elsewhere. The third is \emph{graceful degradation}: the
architecture must specify safe behaviors for every foreseeable
failure mode across its components, ensuring that partial outages
produce conservative, bounded responses rather than unsafe or
unconstrained ones. An architecture that requires all components to
be fully operational in order to behave safely is not suitable for
sovereign infrastructure, where resilience under adversarial or
degraded conditions is a first-order requirement.

\subsection{Layer 1: Physical infrastructure}

At the foundation of the architecture lies the physical
infrastructure examined in Sections~\ref{sec:datacenters}
and~\ref{sec:optical}. AI-oriented data centers host high-density
compute platforms and advanced cooling systems whose design is
governed by the power, thermal, and water constraints analyzed in
Section~\ref{sec:datacenters}. These facilities are interconnected
through metro, regional, and long-haul optical networks whose
latency floors, capacity ceilings, and failure domains define the
feasible geographic scope of sovereign AI operation, as analyzed in
Section~\ref{sec:optical}. Energy systems supply power from grid
connections, on-site renewable generation, and storage assets, often
under tight capacity, carbon intensity, and demand response
obligations~\cite{IEA2023Datacenters,korde2025sovereign}.

The defining characteristic of the physical layer, from an
architectural perspective, is tight cross-domain coupling. A
decision about compute placement determines power draw at the
destination site, which determines cooling load and water
consumption, which may activate a water permit constraint, which
removes that site from the feasible set for future placements. A
change in grid carbon intensity at one site changes the relative
sustainability ranking of available placements, which changes
traffic flows through the optical network, which changes link
utilization and may activate a capacity constraint. These causal
chains propagate across domain boundaries on timescales ranging
from milliseconds for power dynamics to hours for carbon intensity
shifts to seasons for water permit limits. The observability layer
must capture all of them; the control layer must reason across all
of them simultaneously.

A sovereignty-relevant dependency at the physical layer that
deserves explicit architectural attention is the supply chain for
next-generation optical components. Sustaining AI scale requires
continued advances in photonic integration, transceiver density,
and coherent DSP efficiency to improve bandwidth per watt and per
unit of rack space~\cite{Photonics212026,Shekhar2024}. Regions
that lack domestic capacity in photonic component design or
manufacturing face a structural dependency at the physical layer
that propagates upward: even a fully sovereign control architecture
cannot compensate for physical layer constraints imposed by supply
chain disruption or export restriction on critical optical
components. This dependency is analogous to the semiconductor
dependency that has already shaped AI sovereignty policy in
multiple jurisdictions and warrants similar strategic attention.

\subsection{Layer 2: Unified observability}

The observability layer is the architectural mechanism through
which the physical infrastructure becomes a controllable system.
As analyzed in Section~\ref{sec:telemetry}, it ingests streaming
telemetry from four physically distinct domains --- compute and
power, cooling and water, optical network, and grid sustainability
--- each with heterogeneous protocols, schemas, and update rates.
These domains correspond directly to the telemetry structure shown
in Figure~\ref{fig:ref_arch}: compute and power state, optical
KPIs, sustainability signals, and the resulting state fusion
process that assembles these inputs into the unified representation
$\boldsymbol{\theta}(t)$. The layer's functional responsibilities
are schema normalization, timestamp alignment, freshness
validation, confidence scoring, provenance tracking, and
cross-domain fusion, producing a unified infrastructure state
representation $\boldsymbol{\theta}(t)$ that the control layer can
consume without domain-specific knowledge of the underlying
instrumentation.

These functions operate as a processing pipeline rather than
independent steps. Streaming ingestion and schema normalization
align heterogeneous inputs; timestamp alignment enforces a common
temporal reference; freshness certification bounds decision risk;
provenance tracking preserves source traceability; and cross-domain
fusion produces a consistent multi-layer state representation. The
output of this pipeline is a certified, timestamped, and fused
state vector that defines the observable boundary of the system.

The interface contract between the observability layer and the
control layer specifies three properties that are
sovereignty-relevant. First, \emph{completeness}: the state
representation must include all parameters that appear as
constraint inputs in the control layer's optimization, with no
silent gaps that cause the optimizer to evaluate constraints
against unobserved or assumed values. Second, \emph{freshness
certification}: each parameter in the state representation must
carry a timestamp and a freshness flag indicating whether it was
measured within its domain-specific maximum age, estimated from a
forecast within a specified uncertainty bound, or substituted from
a conservative bound because no fresh measurement was available.
The control layer must consume these flags and adjust its behavior
accordingly, holding optimization when critical parameters are
stale rather than proceeding on potentially invalid state. Third,
\emph{provenance}: the origin of each measurement --- the specific
sensor, API endpoint, or estimation model that produced it ---
must be traceable through the state representation to support
audit and fault attribution.

Vendor neutrality at the observability layer is realized through
adherence to open telemetry standards rather than through
vendor-specific management APIs. Optical network telemetry should
be sourced through OpenConfig YANG models and gNMI streaming over
gRPC~\cite{Shakir2018gNMI}. Compute and power telemetry should be
sourced through Redfish~\cite{DMTF2024Redfish} or IPMI interfaces.
Grid sustainability signals should be sourced through open or
documented APIs with explicit freshness and methodology
disclosure~\cite{Cote2025}. Where vendor implementations deviate
from open standards, the observability layer should mediate through
normalization adaptors rather than propagating vendor-specific
representations into the control layer. This architectural boundary
--- open standards at the observability interface, proprietary
implementations below it --- is the mechanism through which the
control layer achieves independence from specific equipment vendors
without requiring that all physical layer components be replaced
with standards-compliant alternatives simultaneously.

\subsection{Layer 3: Coordinated control}

The control layer realizes the agentic architecture described in
Section~\ref{sec:agentic}. It consumes the unified state
representation from the observability layer, applies
domain-specific and coordination-level reasoning to produce a
candidate joint action, and submits that action to the digital
twin for pre-execution validation before releasing it to the
execution layer. As shown in Figure~\ref{fig:ref_arch}, the layer
is structured into three primary components --- the domain agent
tier, the coordination tier, and the digital twin --- complemented
by an LLM interface that supports intent interpretation and
explanation without participating in execution.

The \emph{domain agent tier} comprises specialized agents for
compute placement, power management, cooling control, and optical
routing. Each agent consumes the subset of the state
representation relevant to its domain, applies its policy
specification to generate candidate actions, and exposes those
candidates to the coordination layer through a structured proposal
interface that specifies the action, its expected effect on
relevant state parameters, the constraints it satisfies, and the
constraints it may stress in adjacent domains. This proposal
interface is the mechanism through which the coordination layer
can evaluate joint feasibility without requiring each agent to
model the full cross-domain state.

The \emph{coordination tier} receives proposals from all domain
agents and applies joint feasibility checking to determine whether
a combination of proposals simultaneously satisfies all active
constraints across all domains~\cite{TMF398}. This function is
aligned with cross-domain automation frameworks such as TM Forum
IG1230 and ETSI ZSM, in which policy constraints are enforced as
hard limits rather than optimization penalties. When a jointly
feasible combination exists, it is assembled into a candidate
joint action and forwarded to the digital twin. When no jointly
feasible combination exists, the coordination tier applies a
structured escalation protocol: it identifies the binding
constraint or set of constraints responsible for infeasibility,
generates a human-readable explanation of the conflict, and holds
the system in its current configuration pending operator
instruction. This escalation path is a sovereignty mechanism: it
ensures that constraint violations are never resolved silently by
automated systems making implicit policy decisions, but are always
surfaced to the operator with the information needed to make an
explicit one.

The \emph{digital twin} validates the candidate joint action
against a higher-fidelity model of physical infrastructure than
the optimization model alone can provide~\cite{Cruzes2026}. As
analyzed in Section~\ref{sec:agentic}, this includes physical
dynamics, model abstraction gaps, and cross-domain interactions
that cannot be faithfully encoded as linear constraints in the
optimization formulation. The twin returns either a validation
certificate --- an explicit authorization token required for
execution --- or a rejection specifying which predicted constraint
would be violated, which triggers re-optimization with a
tightened feasible set that excludes the rejected action. The
validation certificate functions as a control invariant: no action
may reach a southbound interface without it. The twin also
continuously reconciles its predictions against measured telemetry
outcomes, flagging model drift when the physical system's response
deviates systematically from prediction and triggering model
recalibration before drift accumulates to the point where
validation becomes unreliable.

A complementary component of the control layer is the \emph{LLM
interface layer}, shown explicitly in Figure~\ref{fig:ref_arch}.
This layer operates strictly in an advisory and interpretive role:
it translates human intent into structured objectives, supports
plan generation, and provides explanations for agent decisions,
but it does not participate in constraint enforcement or action
execution. All LLM-derived outputs must pass through
coordination-layer feasibility checking and digital twin
validation before they can influence system behavior.

\subsection{Layer 4: Safe execution and feedback closure}

The execution layer translates validated actions into
configuration changes applied to physical infrastructure through
domain-specific southbound interfaces. As shown in
Figure~\ref{fig:ref_arch}, these interfaces span four domains:
compute execution (e.g., workload scheduler APIs such as
Kubernetes or Slurm), optical execution (e.g., NETCONF/RESTCONF
interfaces to network controllers), power and cooling control
(e.g., demand response APIs and building management systems), and
audit and traceability systems that record all executed actions.
The layer enforces a strict invariant: only actions that carry a
digital twin validation certificate are submitted to southbound
interfaces. Actions that have not passed pre-execution validation,
or whose validation certificate has expired because infrastructure
state changed between validation and execution, are withheld and
returned to the control layer for re-evaluation against the
current state.

All executed actions generate signed records that bind the action,
the state under which it was validated, and the identity of the
validation process. These records form the audit and traceability
interface shown in Figure~\ref{fig:ref_arch}, enabling post-hoc
verification, compliance auditing, and attribution of
responsibility~\cite{w3ctrace}. This mechanism ensures that
execution remains accountable and that validated control decisions
can be reconstructed and inspected over time.

In sovereign infrastructure deployments, the integrity of the
audit trail itself becomes a sovereignty concern. Signed execution
records stored in a conventional centralized log are vulnerable to
post-hoc modification, selective deletion, or suppression under
regulatory or legal pressure from external actors --- precisely
the scenarios that operational sovereignty is designed to resist.
For deployments where audit integrity under adversarial conditions
is a requirement, execution records can be anchored to a
tamper-resistant trust substrate, such as a permissioned
distributed ledger, to provide non-repudiation, provenance
integrity, and verifiable lifecycle traceability across
infrastructure actions, control decisions, and AI-assisted
operations~\cite{Haque2021Blockchain,Fraga2020Blockchain}.

The architectural role of this mechanism is additive rather than
substitutive. A distributed ledger does not replace the local
control loop, the digital twin validation process, or the signed
execution records described above. It extends them by providing
an independently verifiable commitment to the sequence and content
of sovereignty-relevant events: that a specific action was
proposed at a specific time, validated against a specific twin
state, executed through a specific southbound interface, and
recorded without alteration. This commitment is particularly
relevant in multi-operator or multi-jurisdictional deployments,
where audit records must be accepted as authoritative by parties
who did not participate in their creation and who may have
adversarial interests in contesting their
contents~\cite{Fraga2020Blockchain}.

Permissioned ledger architectures, such as Hyperledger
Fabric~\cite{Androulaki2018Hyperledger}, are better suited to this
role than public blockchain designs in most sovereign
infrastructure contexts, because they allow membership, write
authority, and read access to be governed by the participating
operators and regulators rather than by open consensus. The ledger
does not need to store full execution payloads --- storing
cryptographic hashes of signed records together with sufficient
metadata to identify the action, the validation certificate
reference, and the infrastructure state snapshot is sufficient
for non-repudiation while keeping on-chain data volume
tractable~\cite{Haque2021Blockchain}. This design also decouples
audit integrity from the availability of the primary telemetry and
control systems: even if an operator's internal logging
infrastructure is compromised or unavailable, the on-chain
commitment provides an independent reference point for compliance
auditing and incident reconstruction.

This mechanism is optional in the reference architecture:
deployments that operate within a single administrative domain
with strong internal audit controls and no multi-party
accountability requirement can satisfy the audit and traceability
function entirely through the signed execution records and W3C
PROV-DM~\cite{w3cprov} provenance model described above. The
distributed ledger extension is recommended for deployments that
cross operator boundaries, involve regulated workloads subject to
external audit, or operate in environments where the integrity of
the audit trail may itself be contested.

The feedback path from execution to observability carries two
classes of information that serve distinct purposes, corresponding
to the two feedback loops in Figure~\ref{fig:ref_arch}.
\emph{Feedback~A} (\emph{measured outcomes}) --- the actual power
draw, carbon consumption, water usage, link utilization, and
workload placement observed after execution --- updates the state
representation $\boldsymbol{\theta}(t)$ for the next control
cycle. This is the primary closed-loop signal that confirms
whether executed actions produced their intended effects or
whether corrective action is required.

\emph{Feedback~B} (\emph{residuals}) --- the difference between
digital twin predictions and measured outcomes --- is routed
exclusively to the digital twin for model recalibration. This
signal enables the detection of systematic deviations between
predicted and observed system behavior, triggering model updates
before drift compromises validation reliability. These two
feedback streams must be maintained strictly separate and routed
to their respective consumers; conflating them into a single
monitoring stream removes the model drift detection capability
that is essential for maintaining twin fidelity over time.

\subsection{Graceful degradation under partial failure}

A sovereign architecture must specify safe system behavior for
every foreseeable component failure, not only for nominal
operation. These failure-handling mechanisms preserve operational
sovereignty under degraded conditions by ensuring that control
decisions remain bounded, auditable, and locally governed. Four
failure modes warrant explicit treatment.

When \textbf{telemetry is degraded or unavailable} for one or
more parameters, the observability layer applies the three-tier
freshness policy described in Section~\ref{sec:telemetry}:
forecast substitution with widened uncertainty for slowly varying
parameters, conservative bounding for safety-critical parameters,
and optimization hold for parameters with no safe substitute. The
state representation $\boldsymbol{\theta}(t)$ must reflect these
substitutions explicitly through freshness flags, and the control
layer must restrict its action space accordingly, avoiding
decisions that depend on unverified or stale state.

When the \textbf{digital twin is unavailable}, the control layer
falls back to a conservative operating mode in which only actions
within a pre-specified safe envelope --- defined as actions that
do not increase power draw, do not migrate workloads, and do not
modify routing --- are permitted without validation. This envelope
is defined at deployment time and reviewed periodically; it
represents the operator's judgment about which actions are
sufficiently low-risk to execute without pre-execution simulation.
Actions outside the safe envelope are held until the twin is
restored. The invariant that only validated actions may be
executed is relaxed in a controlled and explicitly bounded way,
preserving safety while maintaining limited operational
continuity.

When the \textbf{optimization is infeasible} --- when no placement
and routing assignment exists that satisfies all active
sustainability, latency, and capacity constraints simultaneously
--- the system interprets infeasibility as a sovereignty signal
rather than a solver failure, as analyzed in
Section~\ref{sec:optical}. Delay-tolerant workloads are deferred;
portable workloads are migrated to the least constrained available
site with an explicit constraint relaxation logged for operator
review; non-portable, latency-sensitive workloads are maintained
at their current placement with a constraint violation alert
escalated to the operator. The system never silently violates a
hard constraint to avoid infeasibility, ensuring that policy
decisions remain explicit and operator-controlled.

When \textbf{agent coordination fails} due to communication loss
between domain agents and the coordination tier, each domain agent
falls back to its most recent valid local policy, freezing its
proposed actions at the last jointly feasible configuration. This
freeze-on-failure behavior prevents domain agents from taking
unilateral actions that may be locally valid but jointly
infeasible, at the cost of reduced responsiveness to
infrastructure changes during the coordination outage. This
mechanism preserves coordinated control semantics by ensuring that
no domain operates outside the last validated global state.

\subsection{Sovereignty assessment framework}

The reference architecture enables a structured assessment of the
sovereignty level achieved by a given deployment, organized across
the four layers and the three design principles. As defined in
Section~\ref{sec:intro}, this assessment must distinguish between
legal sovereignty --- formal control over ownership, jurisdiction,
and access --- and operational sovereignty, which depends on the
ability to observe, decide, validate, and execute actions across
the infrastructure stack. Table~\ref{tab:sovereignty_assessment}
provides a qualitative framework mapping each layer to the
sovereignty properties it contributes, the indicators that signal
high versus compromised sovereignty at that layer, and the primary
risk if those properties are not realized. This framework is
intended as a diagnostic instrument for operators and policymakers
evaluating existing deployments or specifying requirements for new
ones.

\begin{table}[t]
\centering
\caption{Sovereignty assessment framework for AI infrastructure
  deployments. Each architectural layer contributes distinct
  sovereignty properties; compromised sovereignty at any layer
  propagates upward, limiting the effectiveness of sovereign
  control at higher layers regardless of their individual
  integrity.}
\label{tab:sovereignty_assessment}
\begin{tabular}{p{3.0cm} p{3.0cm} p{4.3cm} p{4.2cm}}
\hline
Layer & Sovereignty property & High sovereignty indicator
  & Primary risk if compromised \\
\hline
Physical infrastructure
  & Local energy, cooling, and optical control
  & Domestic grid access; on-site generation; owned fiber routes
  & Environmental limits block deployment; foreign regulatory
    reach over data in transit \\
Unified observability
  & Local, vendor-neutral, governed telemetry
  & Open standards; freshness-certified state; stream-level
    access policy
  & Decisions on stale or externally filtered state;
    telemetry exposure risk \\
Coordinated control
  & Local reasoning, validation, and escalation
  & Locally maintained twin; structured escalation; no external
    control dependency
  & Silent constraint violations; dependence on vendor control
    plane for policy enforcement \\
Safe execution
  & Validated, auditable, reversible actions
  & Validation certificates enforced; residuals tracked;
    southbound interface independence
  & Unvalidated actions on live infrastructure; loss of
    audit trail \\
\hline
\end{tabular}
\end{table}

A deployment achieves high sovereignty when each layer satisfies
its sovereignty properties and the interfaces between layers
enforce the three design principles --- local grounding,
constraint primacy, and graceful degradation --- through explicit
control, validation, and data exchange contracts. Partial
sovereignty, which is the realistic outcome for most deployments
given global supply chain dependencies, is assessed by identifying
which layers and which interfaces fall short of full sovereignty
and how these deficiencies propagate upward to constrain control
at higher layers. This layered assessment enables targeted
investment by linking specific infrastructure gaps to their
operational consequences: strengthening sovereignty at the
physical layer by securing domestic energy and fiber assets, at
the observability layer by adopting open telemetry standards and
governed data sharing, at the control layer by deploying locally
maintained digital twins and structured escalation paths, and at
the execution layer by enforcing validation certificates and
maintaining southbound interface independence from vendor
management platforms.

Table~\ref{tab:implementation_reference} maps each architectural
layer to representative open standards and reference tools that
enable practical implementation. The table abstracts from specific
versions and focuses on the dominant interfaces and frameworks
used in current deployments, including OpenConfig/gNMI for
telemetry~\cite{Shakir2018gNMI}, T-API for topology
abstraction~\cite{TR547}, Redfish for hardware
management~\cite{DMTF2024Redfish}, and OpenTelemetry for
observability~\cite{otel2023}. Streaming platforms such as
Kafka~\cite{raptis2023kafka} and MQTT~\cite{sanjuan2020mqtt} and
carbon-aware interfaces~\cite{Cote2025,carbonawaresdk} complete
the operational stack. Proprietary substitutions reduce vendor
neutrality and must be evaluated for their impact across layers.

\begin{table*}[t]
\centering
\small
\caption{Implementation reference for sovereign AI infrastructure.
  Each layer is mapped to key standards and tools. Proprietary
  substitutions may impact vendor neutrality and sovereignty.
  Tools listed are representative examples of available
  frameworks; listings do not constitute endorsement and should
  be evaluated against current release status before deployment.}
\label{tab:implementation_reference}
\renewcommand{\arraystretch}{1.1}
\begin{tabular}{p{2.3cm} p{3.6cm} p{4.0cm} p{3.7cm}}
\hline
\textbf{Layer} & \textbf{Standards} & \textbf{Tools}
  & \textbf{Sovereignty note} \\
\hline
\multicolumn{4}{l}{\textit{L1 --- Physical}} \\
\hline
Optical transport & OpenConfig; gNMI; T-API
  & GNPy; Telegraf; OpenROADM
  & Prefer domestic fiber; avoid SNMP \\
Compute & Redfish; IPMI & OpenBMC; Redfish Python
  & Prefer Redfish; IPMI lacks freshness \\
Cooling & BACnet; BRICK & Niagara; BRICK toolkit
  & Normalize before $\boldsymbol{\theta}(t)$ \\
Energy & WattTime; Elec.\ Maps; OpenADR & Carbon SDK; PyPSA
  & Freshness $\leq$\,5\,min \\
\hline
\multicolumn{4}{l}{\textit{L2 --- Observability}} \\
\hline
Modeling & YANG; OpenConfig; OTLP & Pyang; OTEL Collector
  & Canonical YANG paths required \\
Transport & Kafka; MQTT & Kafka; Mosquitto
  & Replication $\geq$\,3; TLS \\
Storage & InfluxDB; OpenMetrics; W3C Trace & InfluxDB; Flink
  & Tag source, time, freshness \\
\hline
\multicolumn{4}{l}{\textit{L3 --- Control}} \\
\hline
Optimization & MILP; ACTN; PCE & OR-Tools; HiGHS; GNPy
  & Hard constraints (carbon, water) \\
Coordination & TMF AN; ETSI ZSM
  & Graph-based orchestration frameworks
  & Deterministic escalation \\
Digital twin & CIM; OpenDSS; GNPy & GNPy; OpenDSS; Modelica
  & Continuous reconciliation \\
LLM & OpenAI API; HF & Ollama; vLLM; LlamaIndex
  & Advisory role only \\
\hline
\multicolumn{4}{l}{\textit{L4 --- Execution}} \\
\hline
Compute & K8s; Slurm & Kubernetes; Slurm; Ray
  & Gate on DT validation \\
Network & RESTCONF; NETCONF; gNMI & ONOS; ODL
  & Rollback if OSNR $>$\,1\,dB \\
Power/cooling & OpenADR; BACnet & OpenADR VTN; EcoStruxure
  & Log all actions \\
Audit & OTEL; PROV-DM & Vector; Jaeger; OpenSearch
  & Store signed traces $\geq$\,12\,mo \\
\hline
\end{tabular}
\end{table*}

\section{Sustainability-Aware Automation and Policy Implications}
\label{sec:sustainability_policy}

As AI infrastructure approaches environmental and resource limits,
sustainability can no longer be treated as an external reporting
obligation or a post-hoc optimization step. The preceding sections
have shown that carbon intensity, water availability, and energy
capacity function as hard feasibility constraints that determine
whether AI operation is possible at a given site and time, not as
efficiency metrics to be minimized within an otherwise unconstrained
design space. This section examines the operational and policy
implications of treating sustainability as a first-class constraint
embedded in the closed-loop control architecture developed in
Section~\ref{sec:reference_architecture}. It argues that
sustainability-aware automation is not merely a technical enhancement
but a governance mechanism that reshapes the relationship between
infrastructure operators, regulators, and the physical environment,
and that realizing its full potential requires policy frameworks
designed for dynamic, telemetry-grounded enforcement rather than
static, retrospective compliance.

\subsection{Embedding sustainability in operational control}

\begin{figure*}[t]
  \centering
  \includegraphics[width=\linewidth]{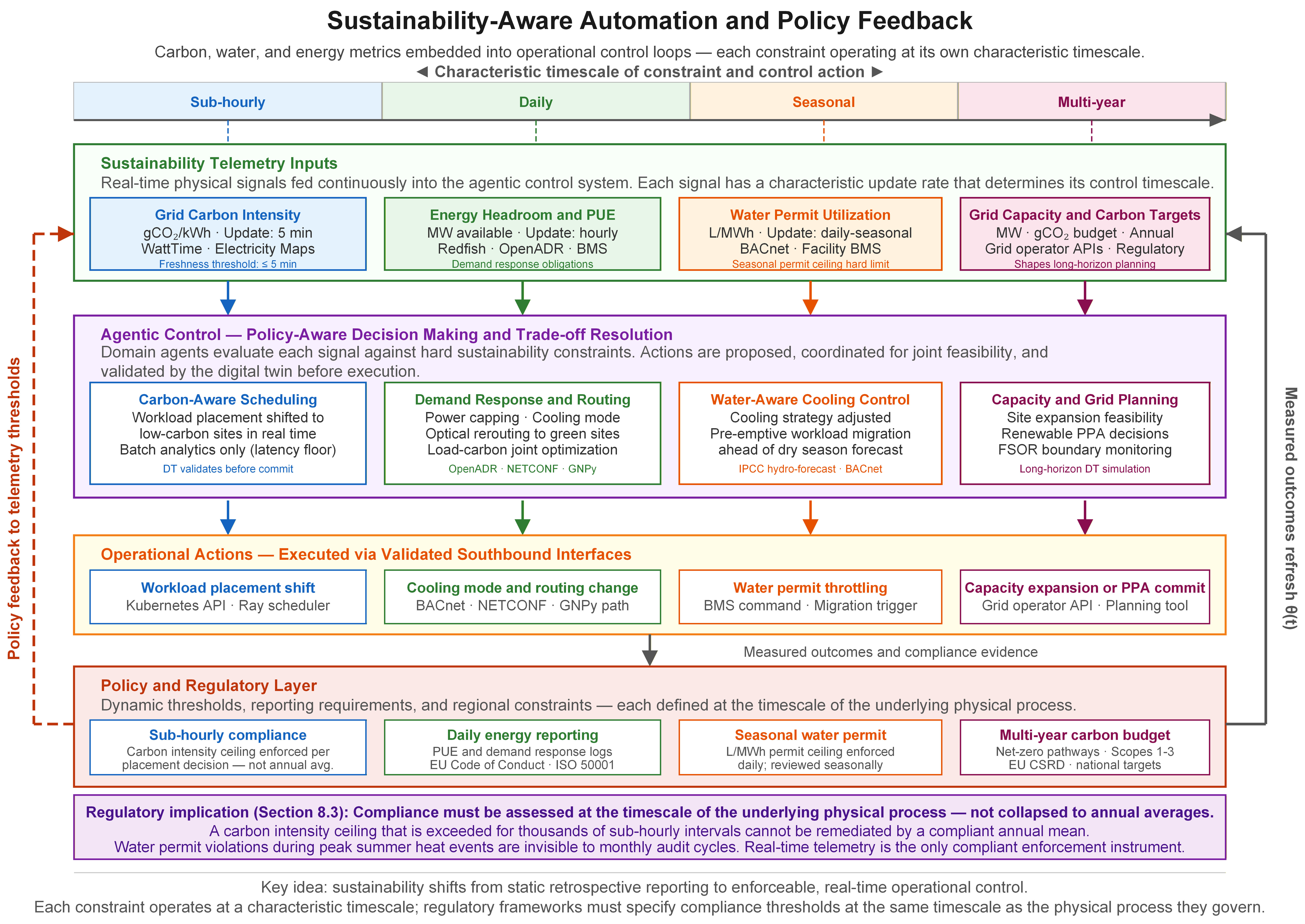}
  \caption{%
    \textbf{Sustainability-aware automation and policy feedback,
    organized by characteristic timescale of constraint and
    control action.}
    The horizontal axis spans four timescale zones ---
    sub-hourly, daily, seasonal, and multi-year --- reflecting
    the characteristic update rate of each sustainability signal
    and the frequency at which corresponding control actions must
    operate.
    Row~1 shows the telemetry inputs feeding each zone: grid
    carbon intensity (5-min
    updates)~\cite{Cote2025}, energy headroom and PUE
    (hourly)~\cite{IEA2023Datacenters}, water permit utilization
    (daily-seasonal)~\cite{YanezBarnuevo2025,IPCC2021WG1}, and
    multi-year grid capacity and carbon
    budgets~\cite{Xiao2025}.
    Row~2 shows the corresponding agentic control actions:
    carbon-aware workload scheduling, demand response and optical
    rerouting, water-aware cooling control with pre-emptive
    migration, and long-horizon capacity planning including FSOR
    boundary monitoring.
    Row~3 shows the executed operational actions through validated
    southbound interfaces.
    Row~4 shows the policy and regulatory layer, with compliance
    thresholds specified at the same timescale as the physical
    process they govern.
    Left and right feedback arrows close the loop: policy
    thresholds are fed back to update telemetry freshness
    requirements; measured outcomes refresh
    $\boldsymbol{\theta}(t)$ in the observability layer.
    The purple box states the central regulatory implication:
    compliance must be assessed at the timescale of the physical
    process, not collapsed to annual averages.%
  }
  \label{fig:sustainability_automation}
\end{figure*}

Sustainability-aware automation integrates carbon intensity, water
usage, and energy availability into the same control loops that
govern performance and reliability, ensuring that decisions reflect
physical and environmental constraints in real time rather than
against abstract annual
targets~\cite{IEA2023Datacenters,radovanovic2021carbonaware,
SILVA2024114019,Xiao2025}. In this sense, sustainability-aware
automation is not merely a technical enhancement but a governance
mechanism that enables operational sovereignty over environmental
constraints. Figure~\ref{fig:sustainability_automation} makes this
structure explicit by organizing the automation framework along a
horizontal timescale axis. The figure should be read from top to
bottom as a closed control loop: telemetry inputs at different
characteristic timescales feed agentic control decisions, which
produce validated operational actions, whose measured outcomes both
update system state and provide compliance evidence to the policy
layer. The figure reveals a constraint that is easy to state but
frequently ignored in regulatory design: each sustainability signal
operates at a characteristic timescale determined by the physical
process it measures, and compliance must be assessed and enforced
at that same timescale rather than collapsed to annual averages.

The four telemetry signals that feed the control system operate at
distinct timescales, each determined by the physical process it
represents. Grid carbon intensity --- measured in grams of
CO\textsubscript{2} equivalent per kilowatt-hour
(gCO\textsubscript{2}eq/kWh) --- changes on sub-hourly timescales
driven by generation dispatch transitions~\cite{Cote2025}.
Commercial providers such as WattTime and Electricity Maps expose
these signals through APIs with update intervals of approximately
five minutes; the architecture enforces a \emph{freshness
threshold} --- a maximum allowable age for a telemetry reading
before it is treated as stale and the dependent decision is held
--- of no more than five minutes for carbon intensity signals,
reflecting the rate at which grid dispatch can shift a site across
a regulatory carbon threshold. Energy headroom and power usage
effectiveness (PUE) are sourced at hourly granularity through
hardware management interfaces such as
Redfish~\cite{DMTF2024Redfish} and demand response platforms based
on OpenADR~2.0b (Open Automated Demand
Response)~\cite{IEA2023Datacenters}, a standard protocol that
enables grid operators to signal power reduction requests to
facility management systems (BMS, building management systems).
Water permit utilization, measured in liters per megawatt-hour
(L/MWh), is reported by facility BMS platforms using BACnet/IP
(Building Automation and Control Networks over IP, ASHRAE
Standard~135), a standard protocol for building system
integration~\cite{YanezBarnuevo2025,IPCC2021WG1}; this signal
operates on daily-to-seasonal granularity because water permit
ceilings are set and enforced seasonally by local regulators.
Multi-year grid capacity and carbon budget signals are sourced
from grid operator APIs and regulatory filings, and shape
long-horizon planning decisions including site expansion
feasibility and renewable energy procurement.

The agentic control system translates each telemetry signal into
a corresponding control action at the appropriate timescale. At
the sub-hourly level, carbon-aware scheduling shifts delay-tolerant
batch analytics workloads to lower-carbon sites in real time;
latency-sensitive workloads cannot be shifted because the
propagation delay floor imposed by optical physics makes distant
low-carbon sites physically infeasible for those workload
classes~\cite{Cote2025}, as analyzed in
Section~\ref{sec:optical}. Every proposed placement is validated
by the digital twin before commitment to the southbound interface,
ensuring that simulated carbon and performance consequences are
confirmed before live execution. At the daily level, demand
response and optical rerouting agents apply power capping, adjust
cooling modes, and reroute optical traffic to lower-carbon sites
using open control interfaces including OpenADR~2.0b for demand
response and NETCONF~\cite{rfc6241} for optical network
reconfiguration, with path feasibility validated through
GNPy~\cite{gnpy2024}. At the seasonal level, water-aware cooling
control agents adjust cooling strategy and trigger pre-emptive
workload migration ahead of dry season periods informed by
hydrometeorological
forecasts~\cite{IPCC2021WG1,YanezBarnuevo2025}, using BACnet
commands to building management systems. At the multi-year level,
capacity and grid planning agents monitor the Feasible Sovereign
Operating Region (FSOR) --- the intersection of sovereignty,
sustainability, and performance constraints within which the
operator can deploy AI workloads --- and evaluate renewable energy
Power Purchase Agreements (PPAs), which are long-term contracts
between electricity consumers and renewable generation owners that
secure a guaranteed supply of low-carbon power at a fixed
price~\cite{IEA2023Datacenters,Xiao2025}. Long-horizon FSOR
boundary monitoring uses digital twin simulation to project how
tightening constraints will affect the feasible deployment region
over planning horizons of three to ten years.

Executed operational actions reach physical infrastructure through
validated southbound interfaces: workload placement decisions
through the Kubernetes API and Ray~\cite{ray2024}, optical
rerouting through NETCONF~\cite{rfc6241} and GNPy-validated paths,
water permit throttling through BMS commands and workload migration
triggers, and capacity expansion decisions through grid operator
APIs and planning tools.

The policy and regulatory layer specifies compliance thresholds at
the same timescale as the physical process each constraint governs.
Sub-hourly carbon intensity ceilings are enforced per placement
decision rather than against an annual average. Daily energy
reporting obligations are met through PUE logs and demand response
records consistent with the EU Code of Conduct on Data Center
Energy Efficiency~\cite{JRC2025} and ISO~50001. Seasonal water
permit ceilings are enforced daily and reviewed at the seasonal
permit cycle. Multi-year carbon budget obligations align with
net-zero pathways reported under Scope~1, Scope~2, and Scope~3
emissions accounting, as required under the EU Corporate
Sustainability Reporting Directive
(CSRD)~\cite{Mayegun2025}.\footnote{Scope~1 covers direct
emissions from owned sources; Scope~2 covers indirect emissions
from purchased energy; Scope~3 covers all other indirect emissions
in the value chain, including embodied carbon in hardware and
supply chain logistics.}

The operational consequence of this embedding is a qualitative
change in the nature of sustainability compliance. Under
conventional reporting models, an operator measures aggregate
carbon emissions over a quarter, compares them against a declared
target, and adjusts strategy for the following period. Violations
are identified retrospectively, after the environmental impact has
already occurred, and remediation operates on timescales of months.
Under sustainability-aware automation, the control system evaluates
carbon intensity at each telemetry cycle and enforces the
operator's carbon policy as a hard constraint on every placement
and routing decision. A site whose grid intensity exceeds the
policy threshold is removed from the feasible placement set
immediately; a workload that would push water consumption above a
permit limit is blocked before execution; a routing change that
would increase carbon exposure above a declared budget triggers
re-optimization rather than execution. Compliance is not reported
after the fact but enforced within the control loop as an
operational invariant: every placement, routing, and resource
allocation decision must satisfy sustainability constraints at
execution time, or it is rejected before reaching the
infrastructure.

Specific control actions enabled by this embedding include
temporal and spatial workload shifting in response to time-varying
grid carbon intensity~\cite{radovanovic2021carbonaware}, dynamic
adjustment of cooling strategies based on water availability and
ambient wet-bulb
temperature~\cite{YanezBarnuevo2025,SILVA2024114019}, pre-emptive
workload migration ahead of predicted water stress periods informed
by hydrometeorological forecasts~\cite{IPCC2021WG1}, and
anticipatory reservation of routing capacity to low-carbon sites
during periods when carbon intensity at current placement sites is
forecast to rise. Each of these actions becomes available only
when sustainability signals are treated as first-class control
inputs on the same footing as latency and utilization metrics.
When they are treated instead as reporting inputs consumed by a
separate sustainability team, the control system lacks the
information needed to act on them at the timescales that matter.

\subsection{Unavoidable trade-offs and their governance}

Embedding sustainability into operational control makes trade-offs
explicit that were previously implicit and resolves them through
explicit policy specification rather than through the accumulated
effect of independent decisions made across organizational silos.
These trade-offs define the boundary conditions of operational
sovereignty: they determine which objectives can be jointly
enforced by the control system and which require explicit policy
relaxation. Three trade-offs are structurally unavoidable and
warrant explicit treatment in any sustainability-aware control
architecture.

The first is the \emph{efficiency--resilience trade-off}.
Minimizing carbon emissions drives workload concentration at sites
with the cleanest grids, reducing the geographic diversity of
active placements and increasing the exposure of the workload
portfolio to local failures, grid instability events, or water
stress periods at those sites. A pure carbon minimization
objective without a resilience constraint will systematically
underinvest in the redundancy needed to maintain service
continuity under adverse conditions. The governance response is to
encode a minimum diversity constraint alongside the carbon
objective: the optimization must maintain feasible alternative
placements for each workload class at sites in distinct failure
domains, even if those alternatives have higher carbon intensity
than the primary placement. The operator's policy specification
determines how much carbon cost is acceptable to purchase a given
level of resilience, making this trade-off explicit and auditable
rather than implicit and opaque.

The second is the \emph{carbon--latency trade-off}. The
green-but-far effect analyzed in Section~\ref{sec:optical} means
that the lowest-carbon sites are frequently not the closest to
demand sources. For latency-sensitive workloads, this trade-off
cannot be resolved by optimization: the propagation delay floor
imposed by optical physics makes some low-carbon placements
physically infeasible regardless of how the objective is weighted.
For delay-tolerant workloads, the trade-off can be navigated
through temporal shifting --- running the workload at a time when
a nearby site has lower carbon intensity --- or through accepting
a higher-latency but lower-carbon placement with explicit
service-level objective relaxation. The policy specification must
define which workload classes are subject to which trade-off
resolution strategies, and the agentic system must enforce those
definitions consistently across all placement decisions.

The third is the \emph{autonomy--scale trade-off}. Prioritizing
local operational sovereignty --- restricting placements to
domestically controlled sites, avoiding cross-border routing, and
minimizing dependence on foreign optical infrastructure --- reduces
the feasible placement set and increases the probability of
infeasibility under tight sustainability constraints. A region
with a carbon-intensive grid and water stress cannot achieve both
maximum sustainability compliance and maximum operational
sovereignty simultaneously if its only compliant sites are abroad.
This trade-off does not have a universal resolution; it reflects
physical and geopolitical constraints that policy must navigate
explicitly rather than dissolve through automation. The agentic
control system's role is to make the trade-off visible and
quantifiable --- specifying precisely which constraints cannot be
satisfied within the sovereignty perimeter and what relaxation
would restore feasibility --- not to resolve it
unilaterally~\cite{IEA2023Datacenters,Xiao2025}.

\begin{figure}[t]
  \centering
  \includegraphics[width=0.82\linewidth]{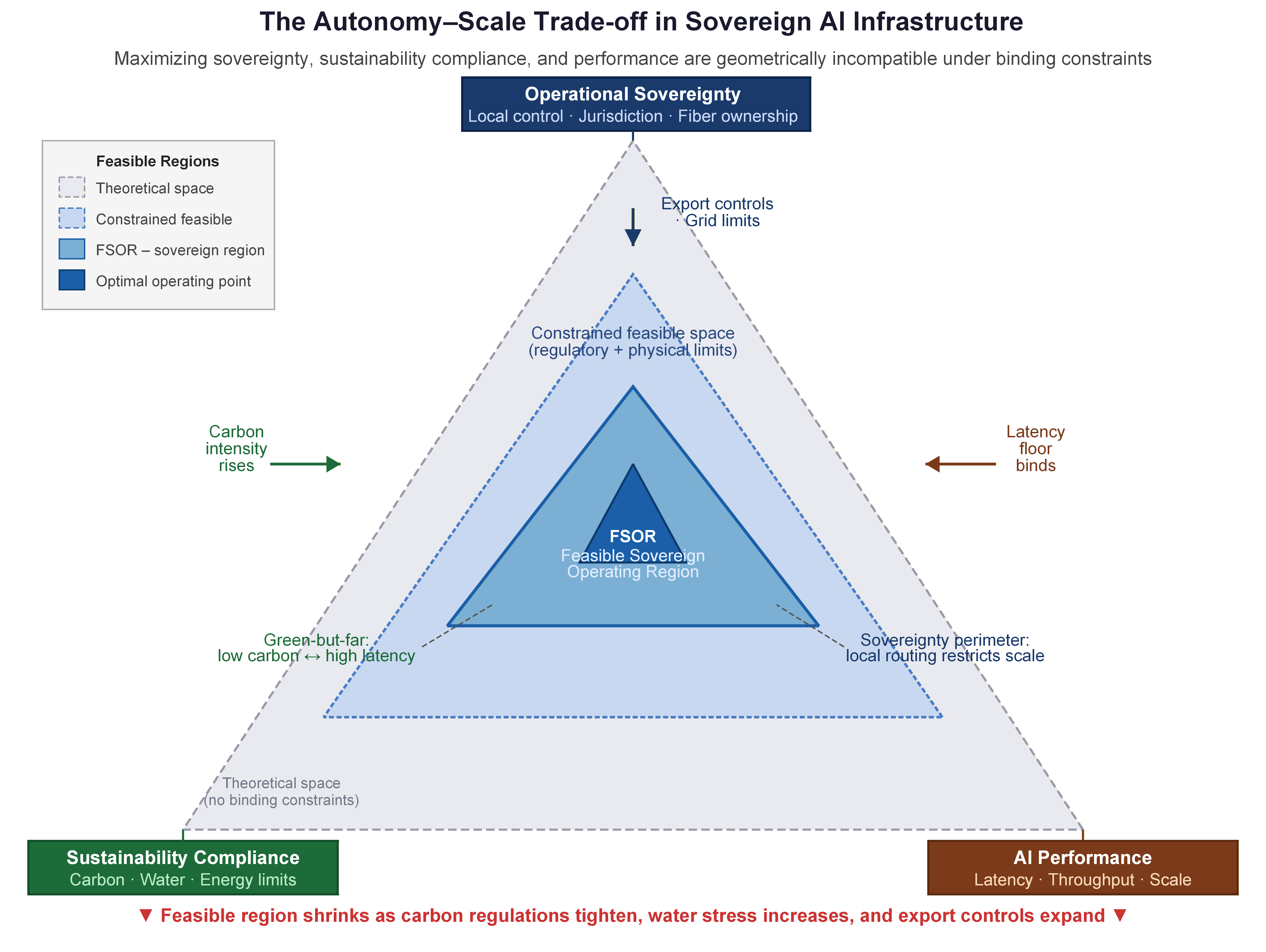}
  \caption{%
    \textbf{The autonomy--scale trade-off in sovereign AI
    infrastructure.}
    Each vertex represents a primary objective: operational
    sovereignty (top), sustainability compliance (bottom left),
    and AI performance (bottom right).
    Nested regions show progressively tighter feasible spaces:
    the theoretical space with no binding constraints (outer,
    dashed grey), the constrained feasible space under regulatory
    and physical limits (middle, dashed blue), and the Feasible
    Sovereign Operating Region --- FSOR --- where all constraints
    are jointly satisfied (inner, solid blue).
    The dark core marks the optimal operating point.
    Constraint pressure from rising carbon intensity, water
    stress, and export controls pushes inward from each edge,
    shrinking the FSOR.
    The green-but-far effect~\cite{IEA2023Datacenters} bounds the
    sustainability--performance edge; sovereignty perimeter
    restrictions bound the sovereignty--performance edge.
    When all three constraints tighten simultaneously, the FSOR
    may contract to the empty set, requiring explicit policy
    relaxation rather than automated resolution.%
  }
  \label{fig:autonomy_scale_tradeoff}
\end{figure}

The three trade-offs described above are not independent
optimization problems but competing pressures on a single bounded
feasible region. Figure~\ref{fig:autonomy_scale_tradeoff}
visualizes this geometry. Each vertex of the triangle represents
a primary objective --- operational sovereignty, sustainability
compliance, and AI performance --- and the nested regions
represent the space of jointly feasible configurations under
progressively tighter constraints. The outermost region is the
theoretical space in which no binding constraints are active. The
intermediate region reflects the constrained feasible space
imposed by regulatory carbon limits, water permit boundaries, and
optical latency floors. The innermost region --- the Feasible
Sovereign Operating Region (FSOR) --- is the intersection where
all three objectives are simultaneously satisfiable under the
operator's declared policy. The central point represents the
optimal operating configuration within that region.

As any single constraint tightens --- through stricter carbon
regulation, seasonal water stress, or expanded export controls on
optical components --- the FSOR shrinks. When constraints from
different vertices tighten simultaneously, the region may contract
to the empty set, at which point no jointly feasible configuration
exists and the system must escalate to human authority for policy
relaxation. The autonomy--scale trade-off is therefore not a
design failure but a structural property of sovereign AI
infrastructure: the geometry of competing constraints makes some
combinations of objectives physically unreachable, and governance
must specify which relaxations are acceptable rather than assume
the trade-off away.

\subsection{Implications for regulators}

Sustainability-aware automation challenges the foundational
assumptions of existing environmental compliance frameworks for
digital infrastructure. Current regulatory models for data center
sustainability are predominantly static and retrospective: they
specify annual energy efficiency targets, require periodic carbon
reporting against declared baselines, and assess compliance
through audit processes that operate on timescales of months to
years~\cite{Mayegun2025}. These mechanisms were designed for
infrastructure whose operating parameters change slowly and whose
environmental impact can be meaningfully characterized by annual
averages. Neither assumption holds for AI infrastructure operating
under the sustainability-aware control architecture described in
this tutorial.

Grid carbon intensity changes on sub-hourly timescales; a facility
that reports a compliant annual average may have operated above
its carbon threshold for thousands of intervals during high-demand
periods when the grid mix was carbon-intensive. Water permit
compliance measured monthly may conceal daily violations during
heat waves. Energy efficiency metrics computed at annual
granularity cannot capture the power dynamics of training
workloads that fluctuate on much shorter timescales. Regulatory
frameworks that rely on these coarse-grained measurements are not
merely imprecise --- they are structurally unable to detect the
compliance failures that matter most in AI infrastructure
environments.

A regulatory framework aligned with sustainability-aware
automation would have four properties. First, it would recognize
real-time telemetry as the authoritative source of compliance
evidence, replacing retrospective self-reporting with continuous,
machine-readable monitoring streams that regulators can access
directly or through certified third-party auditors~\cite{JRC2025}.
Second, it would specify compliance thresholds at the timescale
of the underlying physical process: sub-hourly for carbon
intensity, daily for water permit utilization, and instantaneous
for power capacity constraints, rather than collapsing all
dimensions to annual averages. Third, it would distinguish between
hard regulatory limits --- carbon intensity ceilings, water permit
maxima --- that must be enforced as instantaneous control
constraints, and soft performance targets --- annual average PUE,
lifecycle carbon reduction trajectories --- that can be evaluated
at longer timescales without losing their regulatory meaning.
Fourth, it would treat verifiable, auditable agentic control
actions as compliance evidence: an operator who can demonstrate
that their control system enforced a carbon intensity ceiling on
every placement decision, with a complete action log and digital
twin validation certificate for each, has provided stronger
compliance evidence than one who reports an annual average that
happens to fall below the
threshold~\cite{radovanovic2021carbonaware,JRC2025}.

Designing such a framework requires regulatory capacity that most
jurisdictions currently lack: the technical expertise to evaluate
telemetry pipeline integrity, the legal framework to recognize
automated control actions as binding compliance instruments, and
the institutional processes to audit agentic systems whose
decision logic is distributed across multiple agents and a digital
twin. Building this capacity is a prerequisite for enabling
operational sovereignty over sustainability constraints at the
timescales that matter, and it represents a significant governance
investment that should begin before AI infrastructure scale makes
it urgent.

\subsection{Implications for infrastructure operators}

For infrastructure operators, sustainability-aware automation
transforms sustainability from a compliance function into an
operational capability with direct consequences for cost,
reliability, and competitive
position~\cite{IEA2023Datacenters,SILVA2024114019}. In this
context, sustainability becomes a dimension of operational
sovereignty: the ability to enforce environmental constraints
through control decisions rather than through retrospective
reporting. The transition requires investment across three
dimensions that are interdependent and must be developed in
coordination.

The first dimension is \emph{telemetry infrastructure}. Operators
who lack real-time carbon intensity signals, water permit
monitoring, or sub-minute power draw measurements cannot embed
sustainability into control loops regardless of how sophisticated
their optimization and agentic systems are. The telemetry gaps
identified in Section~\ref{sec:telemetry} --- missing freshness
certification, absence of cross-domain timestamp alignment,
vendor-specific schemas that resist normalization --- are not
merely monitoring deficiencies; they are control limitations that
reduce the precision and reliability of carbon and water
constraint enforcement~\cite{Cruzes2026}.

The second dimension is \emph{control architecture}. Operators
whose sustainability metrics flow into a reporting dashboard rather
than into the objective function and constraint set of their
placement optimizer have not embedded sustainability into control;
they have added a sustainability display to an unchanged control
system. Genuine embedding requires that carbon intensity and water
permit utilization appear in the same mathematical structure as
latency and power capacity constraints, as hard limits that
eliminate infeasible configurations rather than as soft penalties
that can be overridden by performance pressure. Achieving this
requires either redesigning existing orchestration systems or
deploying a sustainability-aware control layer above them that
intercepts and filters placement decisions before execution.

The third dimension is \emph{organizational process}.
Sustainability-aware automation produces decisions at timescales
and volumes that exceed human review capacity: a system managing
hundreds of sites and thousands of workloads may make millions of
sustainability-relevant placement decisions per day. Operators
must develop governance processes that define the scope of
automated authority, specify escalation paths for constraint
conflicts and infeasibility events, establish audit procedures for
reviewing agentic decision logs, and maintain the policy
specifications that bound agent behavior. These processes are as
important as the technical architecture; an agentic system
operating without clear policy specifications and governance
oversight is not an instrument of sovereign control but a source
of unaccountable automation.

\subsection{Implications for hyperscalers}

Hyperscalers occupy a structurally distinct position in the
sustainability governance landscape. Their scale enables capital
investment in advanced telemetry, optimization, and digital twin
infrastructure that is inaccessible to smaller operators, and
their geographic footprint across multiple grid regions enables
sophisticated sustainability arbitrage --- shifting workloads in
real time to the lowest-carbon available sites across a global
infrastructure
portfolio~\cite{radovanovic2021carbonaware,Xiao2025}. At the same
time, their scale amplifies environmental impact: a marginal
efficiency improvement in a hyperscaler's carbon-aware scheduler
affects more aggregate emissions than a comparable improvement in
any individual regional operator's system. Hyperscalers therefore
carry disproportionate responsibility for demonstrating that
sustainability-aware automation works at scale and for
contributing the operational data that can inform the regulatory
frameworks discussed above.

The geopolitical dimension of hyperscaler sustainability practice
is underappreciated. When a hyperscaler's carbon-aware scheduler
shifts workloads away from a high-carbon region, it reduces that
region's AI compute revenue and reinforces the competitive
disadvantage of carbon-intensive grids. When it shifts workloads
toward low-carbon regions, it increases those regions' effective
AI capacity and strengthens their position in the infrastructure
sovereignty landscape. These allocation effects are not neutral:
they reflect and reinforce the structural inequality between
regions with clean energy access and regions without it, examined
in Section~\ref{sec:datacenters}. Regulators in carbon-intensive
regions may find that hyperscaler carbon-aware scheduling
exacerbates their AI infrastructure disadvantage, creating
incentives to regulate or restrict sustainability-driven workload
shifting in ways that conflict with global emissions reduction
objectives.

This tension between national AI sovereignty and global
sustainability optimization reflects a mismatch between
geographically bounded governance and globally optimized control
systems, and it has no clean technical resolution. It therefore
requires explicit policy negotiation at the international
level~\cite{Bakare2024,IEA2023Datacenters}. Such negotiation is
likely to center on mechanisms that align global efficiency with
national sovereignty objectives, including carbon accounting and
credit frameworks that recognize cross-border workload shifting,
minimum domestic capacity requirements for critical AI functions,
and technology transfer or infrastructure co-investment
obligations tied to hyperscaler deployment. These instruments do
not eliminate the tension between sovereignty and sustainability,
but they define how it can be governed in practice, enabling both
objectives to become operational rather than mutually exclusive.

\subsection{Sustainability, sovereignty, and a forward policy
agenda}

The analysis in this tutorial converges on a conclusion that
connects the sustainability and sovereignty dimensions of AI
infrastructure in a way that neither literature has fully
articulated. Sustainability constraints are sovereignty
constraints. A region whose grid is carbon-intensive, whose water
resources are stressed, or whose energy infrastructure cannot
support AI-scale power demand faces sovereignty limits that are
physical rather than legal or political in nature. No regulatory
framework, no data localization requirement, and no model
ownership policy can substitute for the physical capacity to
power, cool, and connect AI infrastructure within locally defined
environmental limits. Conversely, a region that achieves effective
sustainability-aware control of its AI infrastructure --- enforcing
carbon and water limits in real time, shifting workloads in
response to grid signals, and maintaining digital twin models that
make infrastructure behavior verifiable --- demonstrates
operational sovereignty: control authority grounded in the ability
to observe, decide, validate, and act within physical constraints
rather than in formal ownership alone.

This convergence implies a forward policy agenda organized around
four priorities. First, \emph{grid decarbonization as AI
sovereignty policy}: investments in renewable energy capacity,
transmission infrastructure, and grid flexibility are not merely
environmental measures but strategic enablers of sovereign AI
deployment that expand the Feasible Sovereign Operating Region
(FSOR) for regions currently constrained by carbon-intensive
generation. Second, \emph{telemetry standardization as regulatory
infrastructure}: open, interoperable telemetry standards for data
center power, cooling, water, and network state are prerequisites
for both sustainability-aware automation and dynamic regulatory
compliance; their development should be treated as public
infrastructure rather than left solely to vendor-driven
standardization processes. Third, \emph{dynamic compliance
frameworks}: regulatory agencies should develop the technical and
legal capacity to recognize real-time telemetry and auditable
agentic control actions as compliance evidence, replacing annual
reporting cycles with continuous, machine-readable compliance
streams that reflect actual operational behavior. Fourth,
\emph{sovereignty-aware sustainability policy}: international
negotiations on AI infrastructure sustainability should account
for the differential impact of global carbon optimization on
regional AI sovereignty, developing mechanisms --- such as clean
energy investment obligations, technology transfer provisions, or
carbon credit frameworks tied to AI infrastructure deployment ---
that allow sustainability objectives to be pursued without
systematically disadvantaging regions whose grid decarbonization
is incomplete.

Together, these priorities define a governance agenda in which
sustainability and sovereignty are pursued as mutually reinforcing
rather than competing objectives, grounded in the physical
realities of AI infrastructure and implemented through the
telemetry-driven, agentic control architecture described in this
tutorial. In this framework, sovereignty is not asserted through
ownership alone but realized through continuous, validated control
of infrastructure behavior under constraint.

\section{Conclusion}
\label{sec:conclusion}

The rapid expansion of artificial intelligence has shifted the
meaning of sovereignty from a question of data and algorithms to
a question of infrastructure. As AI systems scale, their
feasibility and impact are increasingly determined by access to
power, cooling, water, and optical connectivity, together with
the operational capability to manage these resources under
environmental and geopolitical constraints. AI sovereignty is
therefore not an abstract policy objective, but a property that
emerges from the physical and operational structure of
infrastructure systems.

This tutorial has presented a cross-layer view of AI sovereignty
grounded in engineering practice. It has shown that AI-oriented
data centers, optical transport networks, and energy systems form
a tightly coupled substrate whose constraints must be satisfied
simultaneously. Sustainability limits, particularly those related
to energy, carbon intensity, and water usage, act as hard
boundaries that define where AI can be deployed. Optical networks,
in turn, determine how far and how reliably these capabilities can
be extended across regions. Sovereignty thus emerges from the
joint feasibility of these layers, rather than from any single
dimension in isolation. The Feasible Sovereign Operating Region
(FSOR) formalizes this emergence: it is the set of workload
configurations a given infrastructure can actually sustain under
its physical, environmental, and regulatory endowment, computable
from real-time telemetry and directly actionable as a planning
instrument.

A central contribution of this work is to position telemetry and
agentic AI as the operational foundation of sovereign
infrastructure. Telemetry transforms infrastructure into a
continuously observable system, while agent-based control and
digital twins enable coordinated and validated action across
compute, network, and energy domains. This closed-loop capability
allows operators to enforce constraints, manage trade-offs, and
adapt to changing conditions in real time, making sovereignty a
function of visibility, decision authority, and execution
capability.

These insights have direct implications for both engineering and
policy. For the engineering community, AI system design must move
beyond siloed optimization toward integrated, cross-layer
architectures in which performance, efficiency, sustainability,
and control are jointly considered. Real-time observability and
closed-loop control are no longer optional features, but
fundamental design requirements for operating AI systems at scale
within physical limits. The green-but-far effect --- the
structural tension between sustainability eligibility and latency
admissibility that makes joint placement--routing optimization a
necessity rather than a convenience --- illustrates why
cross-layer integration cannot be deferred to post-design
optimization.

For policymakers and regulators, the results highlight the need
to align governance frameworks with operational reality. Static
notions of compliance based on data location or ownership are
insufficient in systems constrained by dynamic physical
conditions. Effective sovereignty requires mechanisms that enable
verifiable, real-time control over infrastructure behavior,
allowing environmental targets, reliability requirements, and
autonomy objectives to be enforced in practice rather than
declared in principle.

More broadly, this work reframes AI infrastructure as a
feasibility-constrained system in which physical limits,
environmental boundaries, and control capabilities jointly
determine what can be built and operated. In this context,
sovereignty is not defined by isolation, but by the ability to
exercise effective control within these constraints. The future
of AI will therefore be shaped as much by power systems, cooling
technologies, and optical networks as by models and algorithms.
Understanding, designing, and governing these foundations is
essential for building AI ecosystems that are sustainable,
resilient, and operationally sovereign.

\bibliographystyle{elsarticle-num}

\bibliography{biblio}

\end{document}